\documentclass[journal,onecolumn]{IEEEtran}

\usepackage{amsmath,amsthm}
\usepackage{booktabs}
\usepackage{latexsym}
\usepackage{mathrsfs}
\usepackage{amsthm}
\usepackage{amssymb}
\usepackage{amsfonts}
\usepackage{amsbsy}
\usepackage{tikz}
\usepackage{cases}

\usepackage[shortlabels]{enumitem}
\usepackage{url}
\usepackage{array}
\usepackage{pdflscape}
\usepackage{xcolor}
\usepackage{stmaryrd}
\usepackage{verbatim}
\usepackage{braket}
\usepackage{bbm}
\usepackage{bm}
\usepackage{rotating}

\theoremstyle{plain}
\newtheorem{thm}{Theorem}
\newtheorem{lem}{Lemma}

\theoremstyle{definition}
\newtheorem{defn}{Definition}
\newtheorem{example}{Example}
\newtheorem{remark}{Remark}

\usepackage[colorlinks=true]{hyperref}

\allowdisplaybreaks[4]

\begin{document}
\title{Optimal Quantum $(r,\delta)$-Locally Repairable Codes From Matrix-Product Codes}
\author{Meng Cao, and Kun Zhou*
\thanks{Meng Cao and K. Zhou are with Beijing Institute of Mathematical Sciences and Applications, Beijing 101408, China, e-mails: mengcaomath@126.com (M. Cao),  kzhou@bimsa.cn (K. Zhou).}
\thanks{$^*$Corresponding author.}
\thanks{M. Cao is supported by the National Natural Science
Foundation of China under Grant No. 12401684. K. Zhou is supported by the National Natural Science Foundation of China under Grant No. 12401040.}}

\maketitle

\begin{abstract}
This paper studies optimal quantum $(r,\delta)$-LRCs from matrix-product (MP) codes.
We establish a necessary and sufficient condition for an MP code to be an optimal $(r,\delta)$-LRC.
Based on this, we present a characterization for optimal quantum $(r,\delta)$-LRCs from MP codes with nested constituent codes, and also study optimal quantum $(r,\delta)$-LRCs constructed from MP codes with non-nested constituent codes.
Through Hermitian dual-containing and Euclidean dual-containing MP codes, we present five infinite families of optimal quantum $(r,\delta)$-LRCs with flexible parameters.
\end{abstract}


\section{Introduction}\label{sec:intro}

A {\it linear code} over $\mathbb{F}_{q}$ of length $n$, dimension $k$ and minimum (Hamming) distance $d$, denoted as $[n,k,d]_{q}$, is a $k$-dimensional $\mathbb{F}_{q}$-subspace of the vector space $\mathbb{F}_{q}^{n}$.
According to the {\it Singleton bound}, the minimum distance satisfies $d\leq n-k+1$, and if $d=n-k+1$, the code is called a {\it maximum distance separable (MDS) code}.

Classical locally repairable codes (LRCs) are essential for efficient data recovery in distributed storage systems. Optimal LRCs achieve the theoretical limit described by the Singleton-like bound, which establishes the theoretical limit for balancing redundancy and fault tolerance. Since their formalization in foundational works
\cite{Prakash2012} and \cite{Gopalan2012}, these codes have attracted significant research attention due to their practical value in large-scale storage applications.

The quantum counterpart emerged with Golowich and Guruswami's pioneering quantum LRC framework \cite{Golowich2023}, opening new possibilities for quantum data storage. Subsequent research established fundamental bounds and construction methods \cite{Luo2025, Sharma2025}, while Galindo et al. \cite{Galindo2024} extended these to quantum $(r,\delta)$-LRCs. They established equivalence between classical and quantum $(r,\delta)$-LRCs under specific conditions when using Hermitian dual-containing or Euclidean dual-containing codes.

Matrix-product (MP) codes, introduced in \cite{Blackmore2001Matrix}, can produce long classical codes by combining short constituent codes $\mathcal{C}_1,\ldots,\mathcal{C}_N$ with a defining matrix $A\in \mathbb{F}^{N\times r}$. While MP codes efficiently build classical optimal $(r,\delta)$-LRCs (see, e.g., \cite{Luo2022,Galindo2023}), adapting them to quantum optimal $(r,\delta)$-LRCs faced two critical difficult points:
\begin{itemize}
\item  Constituent codes and matrices selection: No systematic/general method existed for choosing suitable constituent codes or defining matrices;

\item Overly Restrictive Conditions: Prior approaches imposed impractical constraints.
\end{itemize}

This paper aims to study optimal quantum $(r,\delta)$-LRCs from matrix-product (MP) codes. The main contributions in this paper are summarized as follows:
\begin{itemize}
\item We establish a necessary and sufficient condition for an MP code to be an optimal $(r,\delta)$-LRC.

\item We characterize optimal quantum $(r,\delta)$-LRCs from MP codes with nested constituent code.

\item We study optimal quantum $(r,\delta)$-LRCs constructed from MP codes with non-nested constituent codes.

\item We construct three infinite families of optimal $(r,\delta)$-LRCs involving Hermitian or Euclidean duality relation.

\item Through MP codes, we present five infinite families of optimal quantum $(r,\delta)$-LRCs with flexible parameters.
\end{itemize}

This paper is organized as follows. Section II reviews fundamental prerequisites, including matrix-product codes, optimal quantum $(r,\delta)$-LRCs. Section III characterizes optimal quantum $(r,\delta)$-LRCs from matrix-product (MP) codes. Section IV constructs three infinite families of optimal $(r,\delta)$-LRCs involving Hermitian or Euclidean duality relation. Section V presents three new infinite families of optimal quantum $(r,\delta)$-LRCs via Hermitian dual-containing MP codes. Section VI presents two new infinite families of optimal quantum $(r,\delta)$-LRCs via Euclidean dual-containing MP codes. Section VII concludes this paper.

\section{Preliminaries}\label{sec:pre}

Let $\mathbb{F}_{q}$ (resp. $\mathbb{F}_{q^{2}}$) denote the finite field with $q$ (resp. $q^{2}$) elements, where $q$ is a prime power. Let $\mathbb{F}_{q}^{N\times M}$ (resp. $\mathbb{F}_{q^{2}}^{N\times M}$) denote the set of all $N\times M$ matrices over $\mathbb{F}_{q}$ (resp. $\mathbb{F}_{q^{2}}$).
For any matrix $A=(a_{i,j})$ over $\mathbb{F}_{q}$, its {\it transpose} is denoted by $A^{\top}=(a_{j,i})$.
For any matrix $B=(b_{i,j})$ over $\mathbb{F}_{q^{2}}$, its {\it conjugate transpose} is denoted by $B^{\dag}=(b_{j,i}^{q})$.

Throughout this paper, we adopt the following conventions and notation.
\begin{itemize}
\item Let $\mathbb{N}=\{0,1,2,\ldots\}$ denote the set of natural numbers (including zero), and let $\mathbb{N}^+=\mathbb{N} \setminus \{0\}$ denote the set of
positive integers.

\item For $n\in \mathbb{N}^+$, define the integer interval $[n]:=\{1,2,\ldots,n\}$.

\item For a finite set $S$, let $|S|$ denote its {\it cardinality}.

\item For a set of vectors $T$, let $\langle T\rangle$ denote the {\it linear span} of $T$.

\item For a square matrix $A$, let $|A|$ denote its {\it determinant}.

\item Let $A$ be an $N\times N$ matrix. Let $1\leq u_{i}\neq u_{j}\leq N$ and $1\leq v_{i}\neq v_{j}\leq N$ for all $1\leq i\neq j\leq s\leq N$. Denote
by\renewcommand{\arraystretch}{0.6}
$\left|A\begin{pmatrix}
u_{1},&u_{2},&\cdots,&u_{s}\\
v_{1},&v_{2},&\cdots,&v_{s}
\end{pmatrix}\right|$ the determinant of the submatrix formed by the $u_{1}'$-th, $u_{2}'$-th, $\ldots$, $u_{s}'$-th rows and the $v_{1}'$-th, $v_{2}'$-th, $\ldots$, $v_{s}'$-th columns of $A$, where $1\leq u_{1}'<u_{2}'<\ldots< u_{s}'\leq N$ and $1\leq v_{1}'<v_{2}'<\ldots< v_{s}'\leq N$ such that
$\{u_{1}',u_{2}',\ldots,u_{s}'\}=\{u_{1},u_{2},\ldots,u_{s}\}$ and $\{v_{1}',v_{2}',\ldots,v_{s}'\}=\{v_{1},v_{2},\ldots,v_{s}\}$.

\item Let ${\rm Sym}_{N}$ denote the set of all permutations of the set $[N]=\{1,2,\ldots,N\}$, known as
the \emph{symmetric group} of order $N$. A permutation $\tau\in{\rm Sym}_{N}$ is called the \emph{identity permutation} if $\tau(i)=i$ for all $i\in[N]$.
For any $1\leq a<b\leq N$, the notation $(a,b)$ represents a \emph{2-cycle} in ${\rm Sym}_{N}$.
Specifically, if $\tau=(a,b)\in{\rm Sym}_{N}$, then $\tau(a)=b$, $\tau(b)=a$ and $\tau(i)=i$ for all $i\in[N]\backslash\{a,b\}$.
\end{itemize}

\subsection{Linear codes}

For a linear code $\mathcal{C}$ of length $n$ over $\mathbb{F}_{q^{2}}$, its {\it Euclidean dual} $\mathcal{C}^{\perp_{\mathrm{E}}}$ and {\it Hermitian dual} $\mathcal{C}^{\perp_{\mathrm{H}}}$ are respectively defined as
\begin{align*}
\mathcal{C}^{\perp_{\mathrm{E}}}=\Big\{\mathbf{x}=(x_{1},\ldots,x_{n})\in\mathbb{F}_{q^{2}}^{n}:\
\langle\mathbf{x},\mathbf{y}\rangle_{\mathrm{E}}=\sum_{i=1}^n x_i y_i=0 \ \mathrm{for} \ \mathrm{all} \  \mathbf{y}=(y_{1},\ldots,y_{n})\in \mathcal{C}\Big\}
\end{align*}
and
\begin{align*}
\mathcal{C}^{\perp_{\mathrm{H}}}=\Big\{\mathbf{x}=(x_{1},\ldots,x_{n})\in\mathbb{F}_{q^{2}}^{n}:\
\langle\mathbf{x},\mathbf{y}\rangle_{\mathrm{H}}=\sum_{i=1}^n x_i y_i^{q}=0 \ \mathrm{for} \ \mathrm{all} \  \mathbf{y}=(y_{1},\ldots,y_{n})\in \mathcal{C}\Big\}.
\end{align*}
In particular, if $\mathcal{C}^{\perp_{\mathrm{E}}}\subseteq\mathcal{C}$ (resp. $\mathcal{C}^{\perp_{\mathrm{H}}}\subseteq\mathcal{C}$), then $\mathcal{C}$ is called
{\it Euclidean} (resp. {\it Hermitian}) {\it dual-containing}.

\subsection{Matrix-product codes}

Matrix-product codes (MP codes, for short), first introduced in \cite{Blackmore2001Matrix} (see also \cite{Ozbudak2002Note}), represent a family of long classical codes formed by combining several classical codes with a defining matrix.
Let $\mathcal{C}_{1},\ldots,\mathcal{C}_{N}$ be linear codes of length $n$ over $\mathbb{F}_{q}$, and let $A=(a_{i,j})_{i\in[N],j\in[M]}\in\mathbb{F}_{q}^{N\times M}$ with $N\leq M$.
The {\it MP code}, denoted by $\mathcal{C}(A):=[\mathcal{C}_{1},\ldots,\mathcal{C}_{N}]\cdot A$, is defined as the set of all matrix-products of the form $[\mathbf{c}_{1},\ldots,\mathbf{c}_{N}]\cdot A$, where $\mathbf{c}_{i}\in\mathcal{C}_{i}$ for $i\in[N]$.
Here, $A$ is referred to as the \emph{defining matrix} of $\mathcal{C}(A)$, and $\mathcal{C}_{1},\ldots,\mathcal{C}_{N}$ are called the \emph{constituent codes} of $\mathcal{C}(A)$.

Every codeword $\mathbf{c}$ in $\mathcal{C}(A)$ can be expressed as
$\mathbf{c}=\big(\sum_{i=1}^N a_{i,1}\mathbf{c}_{i},\sum_{i=1}^N a_{i,2}\mathbf{c}_{i},\ldots,\sum_{i=1}^N a_{i,M}\mathbf{c}_{i}\big)$, where $\mathbf{c}_{i}\in\mathcal{C}_{i}$ for $i\in[N]$. As shown in \cite[Page 480]{Blackmore2001Matrix}, if $G_{i}$ is a generator matrix of $\mathcal{C}_{i}$ for $i\in[N]$, then $\mathcal{C}(A)$ has a generator matrix
\begin{align}\label{generatormatrix}
G:=\left(
\begin{array}{cccc}
a_{1,1}G_{1}& a_{1,2}G_{1}&  \cdots &a_{1,M}G_{1}\\
a_{2,1}G_{2}& a_{2,2}G_{2}& \cdots &a_{2,M}G_{2}\\
\vdots&\vdots&\ddots&\vdots\\
a_{N,1}G_{N}& a_{N,2}G_{N}& \cdots &a_{N,M}G_{N}  \\
\end{array}\right).
\end{align}

\vspace{6pt}

The following two lemmas characterize the parameters of MP codes under different scenarios.

\begin{lem}{\rm (\cite[Page 54]{Ozbudak2002Note})}\label{proposition1}
Let $\mathcal{C}_{i}$ be an $[n,t_{i},d_{i}]_{q}$ linear code for $i\in[N]$, and let $A\in \mathbb{F}_{q}^{N\times M}$ be row full-rank. Then, the MP code $\mathcal{C}(A):=[\mathcal{C}_{1},\ldots,\mathcal{C}_{N}]\cdot A$ is a linear code with parameters
$\big[Mn,\sum_{i=1}^{N}t_{i},d(\mathcal{C}(A))\geq \min\{D_{i}(A)d_{i}:i\in[N]\}\big]_{q}$,
where $D_{i}(A)$ denotes the minimum distance of the linear code on $\mathbb{F}_{q}^{M}$ generated by the first $i$ rows of $A$.
\end{lem}

\begin{lem}{\rm (\cite[Theorem 1]{Hernando2009Construction})}\label{nested-distance}
With the same notation as in Lemma \ref{proposition1}. If $\mathcal{C}_{1},\ldots ,\mathcal{C}_{N}$ are nested, i.e., $\mathcal{C}_{N}\subseteq \ldots \subseteq \mathcal{C}_{1}$, then the minimum distance of $\mathcal{C}(A)$ is $d(\mathcal{C}(A))=\min\{D_{i}(A)d_{i}:i\in[N]\}$.
\end{lem}

Let us now recall an important class of matrices called NSC matrices.

\begin{defn}{\rm (\cite[Definition 3.1]{Blackmore2001Matrix})}\label{defn}
Let $A\in \mathbb{F}_{q}^{N\times M}$ with $N\leq M$. Denote by $A(j_{1},\ldots,j_{i})$ the submatrix formed by the first $i$ rows and the
$j_{1}$-th, $\ldots$, $j_{i}$-th columns of $A$, where $i\in[N]$ and $1\leq j_{1}<\ldots< j_{i}\leq M$. If $A(j_{1},\ldots,j_{i})$ is invertible for all $i\in[N]$ and all ordered $i$-tuples $1\leq j_{1}<\ldots<j_{i}\leq M$, then $A$ is said to be {\it non-singular by columns} ({\it NSC}, for short).
\end{defn}

By the definition of NSC matrices, if $A\in \mathbb{F}_{q}^{N\times M}$ is NSC, then $D_{i}(A)=M-i+1$ for all $i\in[N]$.

\vspace{4pt}

In the following definition, we review the concept of $\tau$-OD matrices, a subclass of NSC matrices introduced in \cite{Cao2024On} and \cite{Cao2025Entanglement}. Examples of $\tau$-OD matrices of small size were provided in \cite{Cao2024On}.

\begin{defn}{\rm (\cite[Sect. 5]{Cao2025Entanglement})}\label{defn10}
Let $\tau\in{\rm Sym}_{N}$. A matrix $A$ is said to be \emph{$\tau$-optimal defining} (\emph{$\tau$-OD}, for short) if it belongs to one of the following two types:
\begin{itemize}
\item [(1)] Type \textrm{I}: $A\in\mathbb{F}_{q}^{N\times N}$ is NSC, and $AA^{\top}$ is a $\tau$-monomial matrix.

\item [(2)] Type \textrm{II}: $A\in\mathbb{F}_{q^{2}}^{N\times N}$ is NSC, and $AA^{\dag}$ is a $\tau$-monomial matrix.
\end{itemize}
\end{defn}

The following lemma presents an infinite family of type II $\tau$-OD matrices, where the permutation $\tau$ is non-idendity.

\begin{lem}{\rm (\cite[Page 12]{Cao2020QIP})}\label{theorem20}
Let $N\mid (q^2-1)$ and $N\nmid(q+1)$. Then, there exists an infinite family of $N\times N$ type II $\tau$-OD matrices over $\mathbb{F}_{q^{2}}$ with
\begin{align}\label{eq-per}
\tau=\begin{pmatrix}
1 & 2 & \cdots  & N\\
t_{0}+1 & t_{1}+1 & \cdots  &  t_{N-1}+1
\end{pmatrix},
\end{align}
where, for each $0\leq i\leq N-1$, the integer $t_{i}$ with $0\leq t_{i}\leq N-1$ is the unique solution to the congruence $i+qt_{i}\equiv 0 \pmod{N}$.
\end{lem}

\vspace{6pt}

The next lemma follows from \cite[Theorem 9]{CaoZhouMP2025}, which reveals the existence of type I (resp. type II) $\tau$-OD matrices for some $\tau\in{\rm Sym}_{N}$,
based on an arbitrary invertible matrix.

\begin{lem}{\rm (\cite[Theorem 9]{CaoZhouMP2025})}\label{LA}
Let $q$ be an odd prime power. Then, the following two statements hold:
\begin{itemize}
\item [(1)] If $A\in\mathbb{F}_{q}^{N\times N}$ is invertible, then there exists a unit lower triangular matrix $L\in\mathbb{F}_{q}^{N\times N}$ such that $LA$ is a
type I $\tau$-OD matrix for some $\tau\in{\rm Sym}_{N}$.

\item [(2)] If $A\in\mathbb{F}_{q^{2}}^{N\times N}$ is invertible, then there exists a unit lower triangular matrix $L\in\mathbb{F}_{q^{2}}^{N\times N}$ such that $LA$
is a type II $\tau$-OD matrix for some $\tau\in{\rm Sym}_{N}$.
\end{itemize}
\end{lem}

The following Lemma \ref{theorem13} is an immediate corollary of \cite[Theorem 12]{CaoZhouMP2025}.

\begin{lem}{\rm (\cite[Theorem 12]{CaoZhouMP2025})}\label{theorem13}
Let $q$ be an odd prime power, and let $\tau\in{\rm Sym}_{N}$. Then, the following two statements hold:
\begin{itemize}
\item [(1)] If $A\in\mathbb{F}_{q}^{N\times N}$ is invertible, then there exists a unit lower triangular matrix $L\in\mathbb{F}_{q}^{N\times N}$ such that
$LAA^{\top}L^{\top}$ is $\tau$-monomial if and only if
\begin{align}\label{eq-t1}
\tau(1)={\rm min}\left\{s\in [N]: n_{1,s}\neq 0\right\},
\end{align}
where $n_{1,s}$ is the $(1,s)$-th entry of $AA^{\top}$, and for all $2\leq i\leq N$,
\begin{align}\label{eq-t2}
\tau(i)={\rm min}\left\{s \in[N]: s\notin\{\tau(1),\ldots,\tau(i-1)\}, \
\left|(AA^{\top})\begin{pmatrix}
1, & \cdots, & i-1, & i \\
\tau(1), & \cdots , & \tau(i-1), & s
\end{pmatrix}\right|\neq 0\right\}.
\end{align}

\item [(2)] If $A\in\mathbb{F}_{q^{2}}^{N\times N}$ is invertible, then there exists a unit lower triangular matrix $L\in\mathbb{F}_{q^{2}}^{N\times N}$ such that
$LAA^{\dag}L^{\dag}$ is $\tau$-monomial if and only if
\begin{align}\label{eq-t11}
\tau(1)={\rm min}\left\{s\in [N]: n_{1,s}\neq 0\right\},
\end{align}
where $n_{1,s}$ is the $(1,s)$-th entry of $AA^{\dag}$, and for all $2\leq i\leq N$,
\begin{align}\label{eq-t12}
\tau(i)={\rm min}\left\{s \in[N]: s\notin\{\tau(1),\ldots,\tau(i-1)\}, \
\left|(AA^{\dag})\begin{pmatrix}
1, & \cdots, & i-1, & i \\
\tau(1), & \cdots , & \tau(i-1), & s
\end{pmatrix}\right|\neq 0\right\}.
\end{align}
\end{itemize}

\end{lem}

\subsection{$(r,\delta)$-LRCs}\label{LRC}

For an $[n, k, d]_q$ linear code $\mathcal{C}$, we say the $i$-th symbol $c_i$ (where $i\in[n]$) of $\mathcal{C}$ has \emph{$(r,\delta)$-locality} if there exists a subset $S_i\subseteq [n]$ containing $i$ and a punctured code $\mathcal{C}|_{S_i}$ such that the length $|S_i|\leq r+\delta-1$ and the distance $d(\mathcal{C}|_{S_i})\geq \delta $. Here, $\mathcal{C}|_{S_i}$ denotes the code $\mathcal{C}$ punctured on the coordinate set $[n]\backslash S_i$ by deleting the components indexed by $[n]\backslash S_i$ in each codeword of $\mathcal{C}$. We call $\mathcal{C}$ an \emph{$(r,\delta)$-LRC} if every symbol has $(r,\delta)$-locality.

In \cite{Prakash2012}, the Singleton-type bound for the parameters $n$, $k$, $d$, $r$ and $\delta$ of an $(r,\delta)$-LRC is established as
\begin{align}\label{rdelta-singleton}
d\leq n-k+1-\bigg(\bigg\lceil \frac{k}{r}\bigg\rceil-1\bigg)(\delta-1).
\end{align}
An $(r,\delta)$-LRC meeting this bound with equality is called an \emph{optimal $(r,\delta)$-LRC}, or simply called optimal.

Suppose $H=(\mathbf{h}_1,\mathbf{h}_2,\ldots,\mathbf{h}_n)$ is an $m\times n$ matrix, where $\mathbf{h}_i$ is a column vector of length $m$ for every $i\in[n]$.
The \emph{support} of $H$ is defined as
\begin{align*}
\text{Supp}(H) :=\{i \in [n]:\; \mathbf{h}_i \neq \mathbf{0}\},
\end{align*}
where $\mathbf{0}$ is the zero column vector of length $m$. Let us recall a useful lemma given in \cite{Luo2022}.

\begin{lem}{\rm (\cite[Lemma 2]{Luo2022})}\label{local}
Let $\mathcal{C}$ be an $[n,k,d]_q$ linear code. Let $r$ and $\delta$ be positive integers, with $\delta > 1$. The $\alpha$-th code symbol of $\mathcal{C}$ has $(r,\delta)$-locality if and only if there exists an $m\times n$ matrix $H$ over $\mathbb{F}_q$ with the following properties:
\begin{itemize}
\item [(1)] The index $\alpha$ is in $\text{Supp}(H)$;

\item [(2)] $|\text{Supp}(H)|\leq  r +\delta-1$ such that any $\delta-1$ nonzero columns of $H$ are linearly independent over $\mathbb{F}_q$;

\item [(3)] $H\mathbf{c}^{\top} = \mathbf{0}$ for every codeword $\mathbf{c}\in \mathcal{C}$.
\end{itemize}
\end{lem}

\subsection{Quantum $(r, \delta)$-LRCs}
As usual, we use the notation $[\mspace{-2mu}[n,k,d]\mspace{-2mu}]_{q}$ to denote a $q$-ary quantum code of length $n$, dimension $q^{k}$, and minimum distance $d$. Such a code is a $q^{k}$-dimensional subspace of the $q^{n}$-dimensional complex Hilbert space $(\mathbb{C}^{q})^{\otimes n} \cong \mathbb{C}^{q^{n}}$. It can detect up to $d-1$ quantum errors and correct up to $\left\lfloor \frac{d-1}{2} \right\rfloor$ quantum errors.

In a recent study \cite{Galindo2024}, Galindo et al. established a fundamental connection between classical codes satisfying the Hermitian (resp. Euclidean) dual-containing property and quantum $(r,\delta)$-LRCs.
Their work effectively characterizes optimal quantum $(r,\delta)$-LRCs induced by classical codes, as formally defined below.

\begin{defn}{\rm (\cite[Definition 31]{Galindo2024})}\label{quan-LRC}
Let $\mathcal{C}\subseteq \mathbb{F}_{q^2}^n$ (resp. $\mathcal{C}\subseteq \mathbb{F}_{q}^n$) be a linear code. Assume that $\mathcal{C}$ is Hermitian (resp. Euclidean) dual-containing, $\dim(\mathcal{C})=\frac{n+k}{2}$ and $\mathcal{C}$ is an $(r, \delta)$-LRC. If $\delta \leq  d(\mathcal{C}^{\perp_{\mathrm{H}}})$ (resp. $\delta \leq  d(\mathcal{C}^{\perp_\mathrm{E}})$) and the induced quantum code with parameters $[\mspace{-2mu}[n,k,\geq d(\mathcal{C})]\mspace{-2mu}]_{q}$ satisfies
\begin{align}
k+2d(\mathcal{C})+2\bigg(\bigg\lceil\frac{n+k}{2r}\bigg\rceil-1\bigg)(\delta-1)=n+2,\label{quan-inqu}
\end{align}
then the induced quantum code with parameters $[\mspace{-2mu}[n,k,d(\mathcal{C})]\mspace{-2mu}]_{q}$ is said to be an \emph{optimal quantum $(r,\delta)$-LRC}.
\end{defn}

\begin{remark}\label{optimal}
In Definition \ref{quan-LRC}, since $k=2\dim(\mathcal{C})-n$, it is straightforward to verify that Eq. \eqref{quan-inqu} is equivalent to
\begin{align*}
d(\mathcal{C})=n-\dim(\mathcal{C})+1-\bigg(\bigg\lceil \frac{\dim(\mathcal{C})}{r}\bigg\rceil-1\bigg)(\delta-1).
\end{align*}
This is equivalent to stating that $\mathcal{C}$ is an optimal $(r,\delta)$-LRC. Therefore, within the framework of Definition \ref{quan-LRC}, an $(r,\delta)$-LRC with Hermitian (resp. Euclidean) dual-containing property is optimal if and only if its induced quantum $(r,\delta)$-LRC is optimal. This conclusion will be applied in the following sections.
\end{remark}

To investigate optimal quantum $(r, \delta)$-LRCs, we need to revisit the following result.

\begin{lem}{\rm (\cite[Theorem 4.1]{ZhouCao2025})}\label{quanum-cl}
Let $\mathcal{C}$ be an optimal $(r,\delta)$-LRC with parameters $[n,k,d]_{q^{2}}$ (resp. $[n,k,d]_{q}$). Then, the following two statements hold:
\begin{itemize}
\item [(1)]
$n-k\geq \lceil \frac{k}{r}\rceil(\delta-1)$, i.e., $d\geq \delta$;

\item [(2)] If $\mathcal{C}$ is Hermitian (resp. Euclidean) dual-containing, then the induced quantum code is an optimal quantum $(r,\delta)$-LRC.
\end{itemize}
\end{lem}

\section{Characterizations of Optimal Quantum $(r,\delta)$-LRCs from Matrix-Product Codes}

This section aims to characterizes optimal quantum $(r,\delta)$-LRCs from matrix-product (MP) codes.

\subsection{Criterion for optimal $(r, \delta)$-LRCs from MP codes}

In the following theorem, we establish a necessary and sufficient condition for an MP code to be an optimal $(r,\delta)$-LRC.

\begin{thm}\label{thm-ess1}
Let $\mathcal{C}(A):=[\mathcal{C}_1,\ldots,\mathcal{C}_{N}]\cdot A$, where $A$ is an $N\times N$ matrix over $\mathbb{F}_{q^{2}}$ (resp. $\mathbb{F}_{q}$), and $\mathcal{C}_i$ is an optimal $(r,\delta)$-LRC with parameters $[n,k_i,d_i]_{q^{2}}$ (resp. $[n,k_i,d_i]_{q}$) for every $i\in [N]$. If $d(\mathcal{C}(A))\leq d_{j_0}$ for some $j_0\in [N]$, and
$\mathcal{C}(A)$ is an $(r,\delta)$-LRC, then $\mathcal{C}(A)$ is an optimal $(r,\delta)$-LRC if and only if the following two conditions hold:
\begin{itemize}
\item [(1)]
$d_i=\delta$ for every $i\in [N]\backslash \{j_0\}$ and $d(\mathcal{C}(A))=d_{j_0}$;

\item [(2)] $\Big\lceil\frac{\sum_{i=1}^N k_i}{r}\Big\rceil=\sum_{i=1}^N\big\lceil \frac{k_i}{r}\big\rceil$.
\end{itemize}
\end{thm}

\begin{IEEEproof}
Define
\begin{align*}
u:=Nn-\sum_{i=1}^N k_i +1-\Bigg(\Bigg\lceil \frac{\sum_{i=1}^N k_i}{r}\Bigg\rceil-1\Bigg)(\delta-1).
\end{align*}

($\Longrightarrow$):
Since $\mathcal{C}(A)$ is an optimal $(r,\delta)$-LRC and $\mathcal{C}_i$ is an optimal $(r,\delta)$-LRC, we have $d(\mathcal{C}(A))=u$, and also derive $d(\mathcal{C}(A))=u\geq d_{j_{0}}$. By the assumption, we derive $d(\mathcal{C}(A))=d_{j_{0}}$.
This implies $n-k_i=\big\lceil \frac{k_i}{r}\big\rceil(\delta-1)$ for every $i\in [N]\backslash \{j_0\}$, and
$\sum_{i=1}^N\big\lceil \frac{k_i}{r}\big\rceil=\Big\lceil \frac{\sum_{i=1}^N k_i}{r}\Big\rceil$. Therefore, for every $i\in [N]\backslash \{j_0\}$, we derive
$d_{i}=\delta$.

\vspace{4pt}

($\Longleftarrow$): In this case, $u=d_{j_{0}}=d(\mathcal{C}(A))$, which implies that $\mathcal{C}(A)$ is an optimal $(r,\delta)$-LRC.

\vspace{4pt}
Therefore, the whole proof is completed.
\end{IEEEproof}

\subsection{Characterization for optimal quantum $(r, \delta)$-LRCs from MP codes with nested constituent codes}

We present the following characterization for optimal $(r,\delta)$-LRCs from MP codes with nested constituent codes.

\begin{thm}\label{thm-comp}
Let $\mathcal{C}(A):=[\mathcal{C}_1,\ldots,\mathcal{C}_{N}]\cdot A$, where $A$ is an $N\times N$ invertible matrix over $\mathbb{F}_{q^2}$ (resp. $\mathbb{F}_{q}$), $\mathcal{C}_i$ is an optimal $(r,\delta)$-LRC with parameters $[n,k_i,d_i]_{q^{2}}$ (resp. $[n,k_i,d_i]_{q}$) for every $i\in[N]$, and $\mathcal{C}_1\supseteq \ldots\supseteq \mathcal{C}_{N}$. Then, the MP code $\mathcal{C}(A)$ is an optimal $(r,\delta)$-LRC if and only if one of the following two cases holds:

Case I: If $D_{N-1}(A)=1$, then
\begin{itemize}
\item [(1)] $\mathcal{C}_1=\ldots=\mathcal{C}_{N}$ and $d_1=\ldots=d_{N}=\delta$;

\item [(2)] $\big\lceil \frac{Nk_{1}}{r}\big\rceil=N\big\lceil \frac{k_1}{r}\big\rceil$.
\end{itemize}

Case II: If $D_{N-1}(A)=2$, then
\begin{itemize}
\item [(1)] $\mathcal{C}_1=\ldots=\mathcal{C}_{N-1}$, $d_1=\ldots=d_{N-1}=\delta$, and $d_N\leq 2\delta $;

\item [(2)] $\big\lceil \frac{(N-1)k_1+k_N}{r}\big\rceil=(N-1)\big\lceil \frac{k_1}{r}\big\rceil+\big\lceil\frac{k_N}{r}\big\rceil$.
\end{itemize}
\end{thm}

\begin{IEEEproof}
By Lemma \ref{nested-distance}, $d(\mathcal{C}(A))\leq d_{N}$.

($\Longrightarrow$): Since $\mathcal{C}(A)$ is an optimal $(r,\delta)$-LRC, we have $d_1=\ldots=d_{N-1}=\delta$ and $d(\mathcal{C}(A))=d_{N}$. Then,
\begin{align}\label{MPdistance}
d(\mathcal{C}(A))=\min\{D_{N-1}(A)\delta,d_{N}\}.
\end{align}
Besides, $D_{N-1}(A)=1$ or $2$.

\vspace{4pt}

{\bf Case I}: If $D_{N-1}(A)=1$, then by Eq. \eqref{MPdistance} and Lemma \ref{quanum-cl}, we have $d(\mathcal{C}(A))=\delta$ and $d_{N}=\delta$.
Note that
\begin{align}
n-k_1=\bigg\lceil \frac{k_1}{r}\bigg\rceil(\delta-1),\label{eq-1th}
\end{align}
and for $1<i\leq N$,
\begin{align}
n-k_i=\bigg\lceil \frac{k_i}{r}\bigg\rceil(\delta-1).\label{eq-ith}
\end{align}

By Eqs. \eqref{eq-1th} and \eqref{eq-ith}, we derive $k_{1}\leq k_{i}$ for $1<i\leq N$. Thus, $k_i=k_1$ for $1<i\leq N$, which implies $\mathcal{C}_1=\ldots=\mathcal{C}_{N}$. By Theorem \ref{thm-ess1}, we obtain $\lceil \frac{Nk_{1}}{r}\rceil=N\lceil \frac{k_1}{r}\rceil$.

\vspace{4pt}

{\bf Case II}: If $D_{N-1}(A)=2$, we have $d_{N}\leq 2\delta$. Applying Eqs. \eqref{eq-1th} and \eqref{eq-ith}, we obtain $k_i=k_1$ for $1<i\leq N-1$, which implies $\mathcal{C}_1=\ldots=\mathcal{C}_{N-1}$.
By Theorem \ref{thm-ess1}, we obtain $\Big\lceil \frac{(N-1)k_1+k_N}{r}\Big\rceil=(N-1)\big\lceil \frac{k_1}{r}\big\rceil+\big\lceil\frac{k_N}{r}\big\rceil$.

\vspace{6pt}

($\Longleftarrow$): We have $d(\mathcal{C}(A))=d_{N}$. By Theorem \ref{thm-ess1}, it follows that $\mathcal{C}(A)$ is an optimal $(r,\delta)$-LRC.

\vspace{4pt}
This completes the whole proof.
\end{IEEEproof}

\begin{remark}\label{rk-compare}
The result presented in \cite[Theorem 5]{Luo2022} provides a sufficient condition for an MP code to be an optimal $(r,\delta)$-LRC. Specifically, \cite[Theorem 5]{Luo2022} states that the MP code $\mathcal{C}(A)=[\mathcal{C}_1,\ldots,\mathcal{C}_{N}]\cdot A$ forms an optimal $(r, \delta)$-LRC when $A$ is an $N\times N$ NSC matrix, $\mathcal{C}_1=\ldots =\mathcal{C}_{N-1}$, $\mathcal{C}_1\supseteq\mathcal{C}_{N}$, and $\Big\lceil \frac{(N-1)k_1+k_N}{r}\Big\rceil=(N-1)\big\lceil \frac{k_1}{r}\big\rceil+\big\lceil\frac{k_N}{r}\big\rceil$. In comparison, Theorem \ref{thm-comp} establishes a necessary and sufficient condition for an MP code to be an optimal $(r,\delta)$-LRC under the condition $\mathcal{C}_1 \supseteq \ldots \supseteq \mathcal{C}_{N}$. Note that in \cite[Theorem 5]{Luo2022}, the defining matrix $A$ of the MP code is required to be NSC, whereas the defining matrix $A$ in Theorem \ref{thm-comp} is not subject to this restriction. From the definition of NSC matrices, we have $D_{N-1}(A) = 2$ for any $N \times N$ NSC matrix $A$. Therefore, Case II in Theorem \ref{thm-comp} generalizes \cite[Theorem 5]{Luo2022}, while Case I in Theorem \ref{thm-comp} presents an entirely new result.
\end{remark}

\begin{thm}\label{Main-thm}
Let $\mathcal{C}(A):=[\mathcal{C}_1, \ldots, \mathcal{C}_{N}]\cdot A$, where $A$ is an $N\times N$ invertible matrix over $\mathbb{F}_{q^2}$ (resp. $\mathbb{F}_{q}$), $\mathcal{C}_i$ is an optimal $(r,\delta)$-LRC with parameters $[n, k_i, d_i]_{q^2}$ (resp. $[n, k_i, d_i]_{q}$) for every $i\in [N]$, and $\mathcal{C}_1\supseteq \ldots\supseteq \mathcal{C}_{N}$. If $\mathcal{C}_{\tau(i)}^{\bot_{\emph{H}}}\subseteq \mathcal{C}_i$ (resp. $\mathcal{C}_{\tau(i)}^{\bot_{\emph{E}}}\subseteq \mathcal{C}_i$) for every $i\in [N]$, where $\tau\in \emph{Sym}_N$ satisfies Eqs. \eqref{eq-t1} and \eqref{eq-t2} (resp. Eqs. \eqref{eq-t11} and \eqref{eq-t12}), then the MP code $\mathcal{C}(A)$ induces an optimal quantum $(r, \delta)$-LRC if and only if one of the following two cases holds:

Case I: If $D_{N-1}(A)=1$, then
\begin{itemize}
\item [(1)] $\mathcal{C}_1=\ldots=\mathcal{C}_{N}$ and $d_1=\ldots=d_{N}=\delta$;

\item [(2)] $\big\lceil\frac{Nk_{1}}{r}\big\rceil=N\big\lceil\frac{k_1}{r}\big\rceil$.
\end{itemize}

Case II: If $D_{N-1}(A)=2$, then
\begin{itemize}
\item [(1)] $\mathcal{C}_1=\ldots=\mathcal{C}_{N-1}$, $d_1=\ldots=d_{N-1}=\delta$, and $d_N\leq 2\delta $;

\item [(2)] $\big\lceil\frac{(N-1)k_1+k_N}{r}\big\rceil=(N-1)\big\lceil\frac{k_1}{r}\big\rceil+\big\lceil\frac{k_N}{r}\big\rceil$.
\end{itemize}

Moreover, when Case I holds, the optimal quantum $(r,\delta)$-LRC induced by $\mathcal{C}(A)$ has parameters
\begin{align}\label{MP-1}
\big[\mspace{-3mu}\big[Nn,2Nk_1-Nn,\delta\big]\mspace{-3mu}\big]_{q}.
\end{align}
When Case II holds, the optimal quantum $(r,\delta)$-LRC induced by $\mathcal{C}(A)$ has parameters
\begin{align}\label{MP-2}
\big[\mspace{-3mu}\big[Nn,2(N-1)k_1+2k_N-Nn,d_N\big]\mspace{-3mu}\big]_{q}.
\end{align}
\end{thm}

\begin{IEEEproof}
($\Longrightarrow$): By Definition \ref{quan-LRC} and Remark \ref{optimal}, $\mathcal{C}(A)$ is an optimal $(r,\delta)$-LRC. By Theorem \ref{thm-comp}, Case I or Case II holds.

\vspace{6pt}

($\Longleftarrow$): Since either Case I or Case II holds, it follows from Theorem \ref{thm-comp} that $\mathcal{C}(A)$ is an optimal $(r,\delta)$-LRC.
Moreover, since $\tau\in \mathrm{Sym}_N$ satisfies Eqs. \eqref{eq-t1} and \eqref{eq-t2} (resp. Eqs. \eqref{eq-t11} and \eqref{eq-t12}), Lemma \ref{theorem13} ensures the existence of a unit lower triangular matrix $L\in\mathbb{F}_{q^{2}}^{N\times N}$ (resp. $L\in\mathbb{F}_{q}^{N\times N}$) such that $LA(LA)^{\dag}$ (resp. $LA(LA)^{\top}$) is a $\tau$-monomial matrix. Define $\mathcal{C}(LA):=[\mathcal{C}_{1},\ldots,\mathcal{C}_{N}]\cdot LA$. Then, $\mathcal{C}(LA)$
is Hermitian (resp. Euclidean) dual-containing and $\mathcal{C}(A)=\mathcal{C}(LA)$. Therefore, $\mathcal{C}(A)$ is Hermitian (resp. Euclidean) dual-containing. According to Definition \ref{quan-LRC} and Remark \ref{optimal}, $\mathcal{C}(A)$ induces an optimal quantum $(r,\delta)$-LRC.

Consequently, $C(A)$ induces an optimal quantum $(r,\delta)$-LRC if and only if either Case I or Case II holds. Let us now prove that
Eq. \eqref{MP-1} (resp. Eq. \eqref{MP-2}) holds when Case I (resp. Case II) holds.

When Case I holds, we have $d(\mathcal{C}(A))=\delta$ and $\dim(\mathcal{C}(A))=Nk_{1}$. By Definition \ref{quan-LRC}, $\mathcal{C}(A)$ induces an optimal quantum $(r,\delta)$-LRC with parameters $\big[\mspace{-3mu}\big[Nn,2Nk_1-Nn,\delta\big]\mspace{-3mu}\big]_{q}$, as presented in Eq. \eqref{MP-1}.

When Case II holds, we have $d(\mathcal{C}(A))=d_{N}$ and $\dim(\mathcal{C}(A))=(N-1)k_{1}+k_{N}$. By Definition \ref{quan-LRC}, $\mathcal{C}(A)$ induces an optimal quantum $(r,\delta)$-LRC with parameters $\big[\mspace{-3mu}\big[Nn,2(N-1)k_1+2k_N-Nn,d_N\big]\mspace{-3mu}\big]_{q}$, as presented in Eq. \eqref{MP-2}.

Therefore, the proof is completed.
\end{IEEEproof}

\subsection{Characterization for optimal quantum $(r,\delta)$-LRCs from MP codes with non-nested constituent codes}

In this section, we study optimal quantum $(r,\delta)$-LRCs constructed from MP codes with non-nested constituent codes.
Compared with prior works \cite{Galindo2023} and \cite{Luo2022} on constructing optimal $(r,\delta)$-LRCs using MP codes, Theorem \ref{thm-ess1} contains  different construction approaches. To illustrate these methodological differences, we first present two representative examples.

\begin{example}\label{esse-exx}
Define an MP code $\mathcal{C}(I_{2}):=[\mathcal{C}_1,\mathcal{C}_2]\cdot I_{2}$, where $I_2$ is the $2\times 2$ identity matrix,
and $\mathcal{C}_1$ and $\mathcal{C}_2$ satisfy:
\begin{itemize}
\item $\mathcal{C}_1\nsubseteq\mathcal{C}_2$ and $\mathcal{C}_2\nsubseteq\mathcal{C}_1$;

\item $\mathcal{C}_1$ and $\mathcal{C}_2$ are MDS codes with identical parameters $[n,k,d]_{q}$, where $d\geq 2$.
\end{itemize}

Then, both $\mathcal{C}_1$ and $\mathcal{C}_2$ are optimal $(k,d)$-LRCs. Let $G_{i}$ (resp. $H_{i}$) be a generator (resp. parity-check) matrix of $\mathcal{C}_i$ for $i\in[2]$. Since $\mathcal{C}(I_{2})$ has a parity-check matrix
\begin{align*}
H:=\begin{pmatrix}
H_{1} & 0\\
0 & H_{2}
\end{pmatrix},
\end{align*}
then $\mathcal{C}(I_{2})$ is an optimal $(k,d)$-LRC.

On the other hand, since $\mathcal{C}(I_{2})$ has a generator matrix
\begin{align*}
G:=\begin{pmatrix}
G_{1} & 0\\
0 & G_{2}
\end{pmatrix},
\end{align*}
we have $d(\mathcal{C}(I_{2}))=d$. By Theorem \ref{thm-ess1}, we derive that $\mathcal{C}(I_{2})$ is an optimal $(k,d)$-LRC.
\end{example}

\begin{remark}
In Example \ref{esse-exx}, since $\mathcal{C}_1$ and $\mathcal{C}_2$ are non-nested, the construction method of the optimal $(k,d)$-LRC $\mathcal{C}(I_{2})$ differs from that of the optimal $(r,\delta)$-LRCs using MP codes in \cite[Theorem 5]{Luo2022}. Furthermore, note that the linear code generated by the defining matrix $I_2$ of the MP code $\mathcal{C}(I_{2})$ is not an $(k,d)$-LRC. This implies that $\mathcal{C}(I_{2})$ does not belong to the class of codes proposed in \cite[Corollary 15]{Galindo2023}, where the $(r,\delta)$-locality is determined by the defining matrix of MP codes. The pairs of non-nested optimal $(k,d)$-LRCs in Example \ref{esse-exx} are abundant. For instance, numerous Reed-Solomon codes $\mathcal{C}_1$ and $\mathcal{C}_2$ satisfy the above conditions, thereby yielding a wide variety of such MP codes.
\end{remark}

\begin{example}\label{esse-ex1}
Let $q$ be an odd prime power. Define an MP code $\mathcal{C}(A):=[\mathcal{C}_1,\mathcal{C}_2]\cdot A$, where
\begin{align*}
A=\begin{pmatrix}
1 & -1\\
1 & 1
\end{pmatrix},
\end{align*}
and $\mathcal{C}_1$ and $\mathcal{C}_2$ are $q$-ary MDS codes that satisfy:
\begin{itemize}
\item $\mathcal{C}_1$ has a generator matrix $G_1=(1,1)$;

\item $\mathcal{C}_2$ has a generator matrix $G_2=(1,-1)$.
\end{itemize}

Then, both $\mathcal{C}_1$ and $\mathcal{C}_2$ are optimal $(1,2)$-LRCs with identical parameters $[2,1,2]_{q}$.
Note that $\mathcal{C}(A)$ has a generator matrix
\begin{align}\label{ex-loc1}
G=\begin{pmatrix}
1 & 1& -1 & -1\\
1 & -1& 1 & -1
\end{pmatrix}.
\end{align}
Then, $d(\mathcal{C}(A))=2$, and the punctured codes $\mathcal{C}(A)|_{\{1,4\}}$ and $\mathcal{C}(A)|_{\{2,3\}}$ ensure that $\mathcal{C}(A)$ is an optimal $(1,2)$-LRC.

On the other hand, since $d(\mathcal{C}(A))=2$, it follows from Theorem \ref{thm-ess1} that $\mathcal{C}(A)$ is an optimal $(1,2)$-LRC.
\end{example}

\begin{remark}
It is not difficult to verify that $\mathcal{C}_1\cap \mathcal{C}_2=\{\mathbf{0}\}$ in Example \ref{esse-ex1}, which implies that $\mathcal{C}_1$ and $\mathcal{C}_2$ are non-nested. Therefore, the construction method of the optimal $(1,2)$-LRC $\mathcal{C}(A)$ in Example \ref{esse-ex1}
differs from that of the optimal $(r,\delta)$-LRCs using MP codes in \cite[Theorem 5]{Luo2022}. Furthermore, note that the linear code generated by the defining matrix $A$ of the MP code $\mathcal{C}(A)$ is not an $(1,2)$-LRC. This implies that $\mathcal{C}(A)$ does not belong to the class of codes proposed in \cite[Corollary 15]{Galindo2023}, where the $(r,\delta)$-locality is determined by the defining matrix of MP codes.
\end{remark}

\vspace{4pt}

\begin{lem}\label{non-handle}
Let $q$ be an odd prime power.
Let $\mathcal{C}_i$ be an $(r,\delta)$-LRC with parameters $[n, k_i, d_i]_q$ for every $i\in[2]$, where $\mathcal{C}_2\subseteq \mathcal{C}_1$.
Let $G_1=(\mathbf{c}_1,\ldots,\mathbf{c}_{n})$ and $G_2=(\mathbf{c}_1',\ldots,\mathbf{c}_{n}')$ be generator matrices of $\mathcal{C}_1$ and $\mathcal{C}_2$, respectively, where $\mathbf{c}_j$ and $\mathbf{c}_j'$ denote the $j$-th columns of $G_1$ and $G_2$ for every $j \in [n]$.
Define a monomially equivalent code $\mathcal{C}_2'$ of $\mathcal{C}_2$, whose generator matrix is $G_2'=(\mathbf{c}_1',-\mathbf{c}_2',\ldots,-\mathbf{c}_n')$. Let
\begin{align}\label{non-mat}
A=\begin{pmatrix}
1 & 1 \\
1 &-1
\end{pmatrix}.
\end{align}
Then, the MP code $\mathcal{C}(A):=[\mathcal{C}_1,\mathcal{C}_2']\cdot A$ is an $(r,\delta)$-LRC.
\end{lem}

\begin{IEEEproof}
By Eq. \eqref{generatormatrix}, we deduce that $\mathcal{C}(A)$ has a generator matrix
\begin{align}\label{min-dist}
\widehat{G}=\begin{pmatrix}
\mathbf{c}_1&\mathbf{c}_2&\cdots&\mathbf{c}_{n}&\mathbf{c}_1&\mathbf{c}_2&\cdots&\mathbf{c}_{n}\\
\mathbf{c}_1'&-\mathbf{c}_2'&\cdots&-\mathbf{c}_n'&-\mathbf{c}_1'&\mathbf{c}_2'&\cdots&\mathbf{c}_n'
\end{pmatrix}.
\end{align}

Define the following two sets:
\begin{align*}
S_1:=\{1\}\cup\{i: n+2\leq i\leq 2n\},\ S_2:=\{i: 2\leq i\leq n+1\}.
\end{align*}
Then, $S_1\cup S_2=[2n]$, and we can verify that the punctured codes $\mathcal{C}(A)|_{S_1}$ and $\mathcal{C}(A)|_{S_2}$ ensure that $\mathcal{C}(A)$ is an $(r,\delta)$-LRC.
\end{IEEEproof}

\begin{thm}\label{MP-ess2}
Let $q$ be an odd prime power. Let $\mathcal{C}_1$, $\mathcal{C}_2$, $\mathcal{C}_2'$, and the matrix $A$ be defined as in Lemma \ref{non-handle},
where $\mathcal{C}_2\subseteq \mathcal{C}_1$. Define an MP code $\mathcal{C}(A):=[\mathcal{C}_1,\mathcal{C}_2']\cdot A$. Then, the following two statements hold:
\begin{itemize}
\item [(1)] $\mathcal{C}_2'\nsubseteq \mathcal{C}_1$ if and only if $\mathcal{C}_2+\mathcal{C}_2'\nsubseteq \mathcal{C}_1$ if and only if the linear code generated by
$G:=(\mathbf{c}_1',\mathbf{0},\ldots,\mathbf{0})$ is not a subcode of $\mathcal{C}_1$.

\vspace{2pt}

\item [(2)] If both $\mathcal{C}_1$ and $\mathcal{C}_2'$ are optimal $(r,\delta)$-LRCs, and $\mathcal{C}_{i}^{\bot_{\mathrm{H}}}\subseteq \mathcal{C}_i$
(resp. $\mathcal{C}_{i}^{\bot_{\mathrm{E}}}\subseteq \mathcal{C}_i$) for every $i\in[2]$, then $\mathcal{C}(A)$ induces an optimal quantum $(r,\delta)$-LRC if and only if the following two conditions hold:
\begin{itemize}
\item $d_1=\delta$ and $d_2\leq 2d_1$;

\item $\lceil \frac{k_1+k_2}{r}\rceil=\lceil \frac{k_1}{r}\rceil +\lceil \frac{k_2}{r}\rceil$.
\end{itemize}
\end{itemize}
\end{thm}

\begin{IEEEproof}
Statement (1) can be directly verified. Let us now prove statement (2). By Lemma \ref{non-handle}, the MP code $\mathcal{C}(A):=[\mathcal{C}_1,\mathcal{C}_2']\cdot A$ is an $(r,\delta)$-LRC. Moreover, we have $d(\mathcal{C}(A))\leq d_2$.
Assume the MP code $\mathcal{C}(A)$ induces an optimal quantum $(r,\delta)$-LRC. By Definition \ref{quan-LRC} and Remark \ref{optimal}, $\mathcal{C}(A)$ is an optimal $(r,\delta)$-LRC. Then, $d_1=\delta$, $\lceil\frac{k_1+k_2}{r}\rceil=\lceil\frac{k_1}{r}\rceil+\lceil\frac{k_2}{r}\rceil$,
and $d(\mathcal{C}(A))=d_{2}$, which implies $d_2\leq 2d_1$. Conversely, assume $d_1=\delta$, $d_2\leq 2d_1$ and
$\lceil \frac{k_1+k_2}{r}\rceil=\lceil \frac{k_1}{r}\rceil +\lceil \frac{k_2}{r}\rceil$. Then, $d(\mathcal{C}(A))=d_2$, and
$\mathcal{C}(A)$ is an optimal $(r,\delta)$-LRC. Since $\mathcal{C}_{2}^{\bot_{\mathrm{H}}}\subseteq \mathcal{C}_2$ (resp. $\mathcal{C}_{2}^{\bot_{\mathrm{E}}}\subseteq \mathcal{C}_2$), we have $(\mathcal{C}_{2}')^{\bot_{\mathrm{H}}}\subseteq \mathcal{C}_{2}'$ (resp. $(\mathcal{C}_{2}')^{\bot_{\mathrm{E}}}\subseteq \mathcal{C}_{2}'$), and $\mathcal{C}(A)$ is Hermitian (resp. Euclidean) dual-containing. According to Definition \ref{quan-LRC} and Remark \ref{optimal}, $\mathcal{C}(A)$ induces an optimal quantum $(r,\delta)$-LRC. This completes the proof of statement (2).
\end{IEEEproof}

\section{Infinite Families of Optimal $(r,\delta)$-LRCs involving the Hermitian or Euclidean duality relation}

In this section, we construct three infinite families of optimal $(r,\delta)$-LRCs involving Hermitian or Euclidean duality relation.

\subsection{The first infinite family of optimal $(r,\delta)$-LRCs involving the Hermitian duality relation}

Let $\mathcal{C}(u,v,t)$ be a linear code over $\mathbb{F}_{q^{2}}$ with parity-check matrix
\begin{align}\label{pari-1}
H=\begin{pmatrix}
A_{1} & & &\\
& A_{2} &  &\\
&  & \ddots &\\
&  &  &A_t\\
B_{1}& B_{2} & \cdots &B_t
\end{pmatrix},
\end{align}
where
\begin{align}\label{qpari-1}
A_i=\begin{pmatrix}
1 & \omega_i & \cdots&  \omega_i^{q-2}\\
\vdots & \vdots & \ddots&  \vdots\\
1 & \omega_i^{u} & \cdots&  \omega_i^{u(q-2)}
\end{pmatrix}\ \mathrm{and} \  B_i=\begin{pmatrix}
1 & \omega_i^{u+1} & \cdots&  \omega_i^{(u+1)(q-2)}\\
\vdots & \vdots & \ddots&  \vdots\\
1 & \omega_i^{u+v} & \cdots&  \omega_i^{(u+v)(q-2)}
\end{pmatrix}
\end{align}
for every $i\in[t]$, with $\omega_1,\ldots,\omega_t\in \mathbb{F}_{q^{2}}$ being primitive $(q-1)$-th roots of unity.

\vspace{6pt}

In the following theorem, we provide our first infinite family of optimal $(r,\delta)$-LRCs,
and show the Hermitian duality relation, i.e., Hermitian dual-containing pairs and Hermitian dual-containing codes, of such codes under different cases.
Compared to the result of \cite[Theorem 5.1]{ZhouCao2025}, Theorem \ref{Opt-1} (3) offers a new structural characterization of these codes.

\begin{thm}\label{Opt-1}
Let $q\geq 5$ be a prime power. Let $u,t\in \mathbb{N}^+$ and $v\in \mathbb{N}$ satisfy $v\leq u$ and $u+v\leq q-2$. Then, the following three statements hold:
\begin{itemize}
\item [(1)] The linear code $\mathcal{C}(u,v,t)$ defined above is an optimal $(r,\delta)$-LRC with parameters
\begin{align*}
[t(q-1),t(q-1)-tu-v,u+v+1]_{q^2},
\end{align*}
where $(r,\delta)=(q-1-u,u+1)$.

\item [(2)] If $u+v\leq \lfloor \frac{q-1}{2}\rfloor -1$, then $\mathcal{C}(u,v,t)$ is Hermitian dual-containing.

\item [(3)] If $2u+v\leq q-2$, then $\mathcal{C}(u,0,t)^{\bot_{\mathrm{H}}}\subseteq \mathcal{C}(u,v,t)$.
\end{itemize}
\end{thm}

\begin{IEEEproof}
Statements (1) and (2) have been established in \cite[Theorem 5.1]{ZhouCao2025}. For the sake of completeness, we provide here a detailed proof of statement (1). The proof of statement (2) can be derived similarly by following our subsequent proof procedure for statement (3).

By Lemma \ref{local}, $\mathcal{C}(u,v,t)$ is an $(q-1-u,u+1)$-LRC.
Let $E=\{\mathbf{v}_i:\;i\in[t(q-1)]\}$ be the set of column vectors of the parity-check matrix $H$ as defined in Eq. \eqref{pari-1}. For any subset $T\subseteq E$ with $|T|=u+v$, we consider its partition
\begin{align*}
T:=\cup_{j=1}^t T_j,\ \mathrm{with}\ T_j\subseteq \{\mathbf{v}_i:\;(j-1)(q-1)+1\leq i\leq j(q-1)\}.
\end{align*}

If $|T_1|\leq u$, then $\langle T_1\rangle \nsubseteq \langle \cup_{i=2}^t T_i\rangle$ and all the column vectors in $\langle T_1\rangle $ are linearly independent.
If $|T_1|\geq u+1$, then $|T_j|\leq u-1$ for every $2\leq j\leq t$. Thus, $\cup_{j=2}^t T_j\nsubseteq \langle T_1\rangle$.
Hence, any $u+v$ columns of $H$ are linearly independent. Obviously, the first $u+v+1$ columns of the parity-check matrix $H$ in Eq. \eqref{pari-1} are linearly dependent. Consequently, $d(\mathcal{C}(u,v,t))=u+v+1$.

Since $r=q-1-u$, $k=t(q-1)-tu-v$ and $u+v\leq q-2$, we can straightforwardly verify that $\lceil \frac{k}{r}\rceil=\lceil \frac{t(q-1-u)-v}{q-1-u}\rceil=t$, and that the minimum distance of $\mathcal{C}(u,v,t)$ achieves equality in the bound \eqref{rdelta-singleton}. Therefore, $\mathcal{C}(u,v,t)$ is an optimal $(r,\delta)$-LRC.
This completes the proof of statement (1).

Let us now prove statement (3). Note that the parity-check matrix $H$ of $\mathcal{C}(u,v,t)$, defined in Eq. \eqref{pari-1}, can be written as
\begin{align*}
H=\begin{pmatrix}
H_{0}\\
B
\end{pmatrix},\ \mathrm{with} \
H_{0}:=\begin{pmatrix}
A_{1}  & $$& $$\\
$$ & \ddots & $$\\
$$ & $$ & A_t
\end{pmatrix}\ \mathrm{and}  \ B:=(B_{1}, \cdots, B_t).
\end{align*}

Let $\mathbf{n}_{i}$ denote the $i$-th row of $H$ for every $i\in[tu+v]$. Then,
\begin{enumerate}[wide, itemsep=0pt, leftmargin =0pt, widest={{\bf Case $1$}}]
\item[i)] For any $i\neq i'\in[t]$ and $j,j'\in[u]$, we directly obtain
\begin{align*}
\langle\mathbf{n}_{(i-1)u+j},\mathbf{n}_{(i'-1)u+j'}\rangle_{\mathrm{H}}=0.
\end{align*}

\item[ii)] For any $i\in[t]$ and $j,j'\in[u]$, we derive
\begin{align*}
\langle\mathbf{n}_{(i-1)u+j},\mathbf{n}_{(i-1)u+j'}\rangle_{\mathrm{H}}=0.
\end{align*}

\item[iii)] For any $i\in [t]$, $j\in [u]$ and $j'\in [v]$, we derive
\begin{align*}
\langle\mathbf{n}_{(i-1)u+j},\mathbf{n}_{tu+j'}\rangle_{\mathrm{H}}=0.
\end{align*}
\end{enumerate}

Hence, $\mathcal{C}(u,0,t)^{\bot_{\mathrm{H}}}\subseteq \mathcal{C}(u,v,t)$, which verifies statement (3). Therefore, the whole proof is completed.
\end{IEEEproof}

\subsection{The second infinite family of optimal $(r,\delta)$-LRCs involving the Hermitian duality relation}

Let $x_{i,j}=\lambda_i \omega_i^j$, $y_a=\zeta^a$ and $z_{i,j}=\mu_i \omega_i^j$ for $i\in[t]$, $0\leq j\leq q-2$ and $0\leq a\leq v-1$, where
\begin{itemize}
\item $\omega_1,\ldots,\omega_t\in \mathbb{F}_{q^{2}}$ are primitive $(q-1)$-th roots of unity;

\item $\zeta \in \mathbb{F}_{q^{2}}$ is a primitive $v$-th root of unity with $v|(q-1)$;

\item $\lambda_i,\mu_i\in \mathbb{F}_{q^{2}}$ satisfy $\{\lambda_i,\mu_i,\frac{\mu_i}{\lambda_i}\}\subseteq\mathbb{F}_{q^{2}}^{\ast}\backslash
\{\omega_i^j: 1\leq j\leq q-1\}$ for $i\in[t]$.
\end{itemize}
Let $\mathcal{C}(s,l,v,t)$ be a linear code over $\mathbb{F}_{q^{2}}$ with parity-check matrix
\begin{align}\label{qpari-2}
H=\begin{pmatrix}
A_{1} & & &\\
& A_{2} &  &\\
&  & \ddots &\\
&  &  &A_t\\
B_{1}& B_{2} & \cdots &B_t
\end{pmatrix},
\end{align}
where
\begin{align}\label{ha}
A_i=
\left (
\begin{array}{cccc cccc cccc}
x_{i,0} & x_{i,1} & \cdots& x_{i,q-2}&y_0 &y_1&\cdots& y_{v-1}& &  & & \\
\vdots & \vdots & \ddots& \vdots&\vdots &\vdots&\ddots& \vdots& &  & & \\
x_{i,0}^s & x_{i,1}^s & \cdots& x_{i,q-2}^s&y_0^s &y_1^s&\cdots& y_{v-1}^s& &  & & \\
& & & &y_0 &y_1&\cdots& y_{v-1}&z_{i,0} & z_{i,1} & \cdots& z_{i,q-2}\\
& & & &\vdots &\vdots&\ddots& \vdots&\vdots & \vdots & \ddots& \vdots\\
& & & &y_0^s &y_1^s&\cdots& y_{v-1}^s&z_{i,0}^s & z_{i,1}^s & \cdots& z_{i,q-2}^s
\end{array}
\right )
\end{align}
and
\begin{align}
B_i=
\left (
\begin{array}{cccc cccc c}
x_{i,0}^{s+1} & \cdots& x_{i,q-2}^{s+1}&y_0^{s+1} & \cdots& y_{v-1}^{s+1}&z_{i,0}^{s+1} & \cdots& z_{i,q-2}^{s+1} \\
\vdots & \ddots& \vdots &\vdots & \ddots& \vdots&\vdots & \ddots& \vdots \\
x_{i,0}^{s+l} & \cdots& x_{i,q-2}^{s+l}&y_0^{s+l} & \cdots& y_{v-1}^{s+l}&z_{i,0}^{s+l} & \cdots& z_{i,q-2}^{s+l}
\end{array}
\right )
\end{align}
for every $i\in[t]$.

\vspace{6pt}

In the following theorem, we construct the second infinite family of optimal $(r,\delta)$-LRCs,
and show the Hermitian duality relation, i.e., Hermitian dual-containing property and Hermitian dual-containing pairs, of such codes under different cases.

\begin{thm}\label{Opt-3}
Let $q\geq 8$ be a prime power. Let $s,v,t\in \mathbb{N}^+$ and $l\in \mathbb{N}$ satisfy
\begin{itemize}
\item  $v|(q-1)$, $s +l \leq v-1$ and $l \leq \lfloor \frac{s}{2}\rfloor +1$;

\item  $vt\leq q+v-s-l-2$.
\end{itemize}
Then, the following three statements hold:
\begin{itemize}
\item [(1)] The linear code $\mathcal{C}(s,l,v,t)$ defined above is an optimal $(r,\delta)$-LRC with parameters
\begin{align*}
[t(2q+v-2),t(2q+v-2)-2st-l,s+l+1]_{q^2},
\end{align*}
where $(r,\delta)=(q+v-s-1,s+1)$.

\item [(2)] If $s+l\leq \lfloor \frac{v}{2}\rfloor-1$, then $\mathcal{C}(s,l,v,t)$ is Hermitian dual-containing.

\item [(3)] If $2s+l\leq v-1$, then $\mathcal{C}(s,0,v,t)^{\bot_{\mathrm{H}}}\subseteq \mathcal{C}(s,l,v,t)$.
\end{itemize}
\end{thm}

\begin{IEEEproof}
By Lemma \ref{local}, $\mathcal{C}$ is an $(q+v-s-1,s+1)$-LRC. Let $E=\{\mathbf{v}_i: i\in[t(2q+v-2)]\}$ denote the set of all column vectors of the parity-check matrix $H$ as defined in Eq. \eqref{qpari-2}. For any subset $T\subseteq E$ with $|T|=s+l$, we consider its partition
\begin{align*}
T:=\cup_{j=1}^t T_j, \ \mathrm{with} \  T_j\subseteq \{\mathbf{v}_i: (j-1)(2q+v-2)+1\leq i\leq j(2q+v-2)\}.
\end{align*}

Without loss of generality, we assume $T_1\neq \emptyset$ and define the partition:
\begin{align*}
T_1:=T_{11}\cup T_{12} \cup T_{13},
\end{align*}
where
\begin{numcases}{}
T_{11}=T_1\cap \{\mathbf{v}_i: i\in[q-1]\}, \\
T_{12}=T_1\cap \{\mathbf{v}_i: q\leq i\leq q+v-1\}, \\
T_{13}=T_1\cap \{\mathbf{v}_i: q+v\leq i\leq 2q+v-2\}.
\end{numcases}

If $|T_{11}|+|T_{12}|\leq s$, then $\langle T_1\rangle \nsubseteq \langle \cup_{i=2}^t T_i\rangle$ and all the column vectors in $\langle T_1\rangle $ are linearly independent. If $|T_{11}|+|T_{12}|\geq s+1$, we discuss the following two cases:
\begin{enumerate}[wide, itemsep=0pt, leftmargin =0pt, widest={{\bf Case $1$}}]
\item[{\bf Case $1$}:] Suppose $|T_{12}|\geq s+1$. Suppose
\begin{align*}
T_{11}=\{\bm{\alpha}_{1},\bm{\alpha}_{2},\ldots,\bm{\alpha}_{f}\},\ T_{12}=\{\bm{\beta}_1,\bm{\beta}_2,\ldots,\bm{\beta}_{s+g}\},\ T_{13}=\{\bm{\gamma}_1,\bm{\gamma}_2,\ldots,\bm{\gamma}_h\},
\end{align*}
where $f,h\in \mathbb{N}$, $g\in \mathbb{N}^{+}$, and $f+s+g+h\leq s+l$, i.e., $f+g+h\leq l$. Here, we adopt $T_{11}=\emptyset$ (resp. $T_{13}=\emptyset$) if $f=0$ (resp. $h=0$).
Since $\{\lambda_i,\mu_i,\frac{\mu_i}{\lambda_i}\}\subseteq  \mathbb{F}_{q^{2}}^{\ast}\backslash \{\omega_i^j: j\in[q-1]\}$, we derive that all the $x_{i,j}$'s, $y_a$'s, and $z_{i,j}$'s are distinct elements. Since $s+g\leq s+l\leq v-1$, we obtain
\begin{align}\label{equ-inde}
\sum_{i=1}^f a_i\bm{\alpha}_i+\sum_{m=1}^{s+g} b_m\bm{\beta}_m+\sum_{u=1}^h c_u\bm{\gamma}_u=\mathbf{0}\Longrightarrow a_i=b_m=c_u=0
\end{align}
for every $i\in[f]$, $m\in[s+g]$ and $u\in[h]$. Therefore, the column vectors in $T_{1}$ are linearly independent.

\vspace{2pt}

\item[{\bf Case $2$}:] Suppose $|T_{12}|\leq s$. In this case, we can set $|T_{11}|=i+m+1$ and $|T_{12}|=s-i$ for some $0\leq i\leq s$ and $0\leq m\leq l-1$.
Since $|T|=s+l$, we have $|T_1|\leq s+l$ and $|T_{13}|\leq l-m-1$. If $|T_{12}|+|T_{13}|\leq s$, then $\langle T_1\rangle \nsubseteq\langle \cup_{i=2}^t T_i\rangle$. Otherwise, we obtain $|T_{11}|+|T_{13}|\leq s$. Combining this with the fact that $x_{1,i}\neq z_{1,j}$ for all $0\leq i,j\leq q-2$, we derive $\langle T_1\rangle \nsubseteq \langle \cup_{i=2}^t T_i\rangle$ and all the column vectors in $\langle T_1\rangle $ are linearly independent.
\end{enumerate}

\vspace{2pt}

Similar to the discussions in {\bf Case $1$} and {\bf Case $2$} above, we deduce that any $s+l$ columns of $H$ are linearly independent.
Observe that the first $s+l+1$ columns of $H$ are linearly dependent. Therefore, we derive $d(\mathcal{C}(s,l,v,t))=s+l+1$.

Note that $r=q+v-s-1$ and $k=t(2q+v-2)-2st-l=2t(q+v-s-1)-(l+vt)$. By the condition $vt\leq q+v-s-l-2$, we have $l+vt\leq q+v-s-2$. Hence, $\lceil\frac{k}{r}\rceil=2t$, and it is easy to verify that the linear code $\mathcal{C}(s,l,v,t)$ is optimal. This completes the proof of statement (1).

\vspace{5pt}

The parity-check matrix $H$ of $\mathcal{C}(s,l,v,t)$, defined in Eq. \eqref{pari-1}, can be written as
\begin{align*}
H=\begin{pmatrix}
H_{0}\\
B
\end{pmatrix},\ \mathrm{with} \
H_{0}:=\begin{pmatrix}
A_{1}  & $$& $$\\
$$ & \ddots & $$\\
$$ & $$ & A_t
\end{pmatrix}\ \mathrm{and}  \ B:=(B_{1}, \cdots, B_t).
\end{align*}

Let $\mathcal{S}_{0}$ (resp. $\mathcal{S}$) denote the set of all rows of $H_{0}$ (resp. $H$).
Following a similar approach to the proof of Theorem \ref{Opt-1} (3), we deduce that when $s+l\leq \lfloor \frac{v}{2}\rfloor-1$, the equality $\langle\mathbf{y}',\mathbf{y}''\rangle_{\mathrm{H}}=0$ holds for any $\mathbf{y}'\in \mathcal{S}$ and $\mathbf{y}''\in \mathcal{S}$. This implies $\mathcal{C}(s,l,v,t)^{\bot_{\mathrm{H}}}\subseteq \mathcal{C}(s,l,v,t)$, thereby proving statement (2).
Similarly, we deduce that when $2s+l\leq v-1$, the equality $\langle\mathbf{x},\mathbf{y}\rangle_{\mathrm{H}}=0$ holds for any $\mathbf{x}\in \mathcal{S}_{0}$ and $\mathbf{y}\in \mathcal{S}$. This implies $\mathcal{C}(s,0,v,t)^{\bot_{\mathrm{H}}}\subseteq \mathcal{C}(s,l,v,t)$, which verifies statement (3).

Therefore, the whole proof is completed.
\end{IEEEproof}

\subsection{The third infinite family of optimal $(r,\delta)$-LRCs involving the Euclidean duality relation}

Let $\mathcal{D} (u,v,t,m)$ be a linear code over $\mathbb{F}_{q}$ with parity-check matrix
\begin{align}
H=\begin{pmatrix}
A_{1} & & &\\
& A_{2} &  &\\
&  & \ddots &\\
&  &  &A_t\\
B_{1}& B_{2} & \cdots &B_t
\end{pmatrix}, \label{pari-1x}
\end{align}
where
\begin{align}
A_i=\begin{pmatrix}
1 & \zeta_i & \cdots&  \zeta_i^{q-2}\\
\vdots & \vdots & \ddots&  \vdots\\
1 & \zeta_i^{u} & \cdots&  \zeta_i^{u(q-2)}
\end{pmatrix}\ \mathrm{and} \  B_i=\begin{pmatrix}
1 & \zeta_i^{u+1} & \cdots&  \zeta_i^{(u+1)(q-2)}\\
\vdots & \vdots & \ddots&  \vdots\\
1 & \zeta_i^{u+v} & \cdots&  \zeta_i^{(u+v)(q-2)}
\end{pmatrix} \label{qpari-1}
\end{align}
for every $i\in[t]$, with $\zeta_1,\ldots,\zeta_t\in \mathbb{F}_{q}$ being primitive $m$-th roots of unity, and $m\mid(q-1)$.

\vspace{6pt}

In the following theorem, we present an infinite family of optimal $(r,\delta)$-LRCs involving the Euclidean dual-containing relation.

\begin{thm}\label{Euclidean}
Let $q\geq 5$  be a prime power. Let $u,t,m\in \mathbb{N}^+$ and $v\in \mathbb{N}$ satisfy $v\leq u$, $u+v\leq m-1$ and $m\mid(q-1)$.
Then, the following three statements hold:
\begin{itemize}
\item [(1)] The linear code $\mathcal{D}(u,v,t,m)$ defined above is an optimal $(r,\delta)$-LRC with parameters
\begin{align*}
[tm,tm-tu-v,u+v+1]_{q},
\end{align*}
where $(r,\delta)=(m-u,u+1)$.

\item [(2)] If $u+v\leq \lfloor \frac{m}{2}\rfloor -1$, then $\mathcal{D}(u,v,t,m)$ is Euclidean dual-containing.

\item [(3)] If $2u+v\leq m-1$, then $\mathcal{D}(u,0,t,m)^{\bot_{\mathrm{E}}}\subseteq \mathcal{D}(u,v,t,m)$.
\end{itemize}
\end{thm}

\begin{IEEEproof}
Using a similar proof technique as in Theorems \ref{Opt-1} and \ref{Opt-3}, we can derive the required results. The detailed proof is therefore omitted here for conciseness.
\end{IEEEproof}

\begin{remark}
Note that the $q$-ary optimal $(r,\delta)$-LRCs in Theorem \ref{Euclidean} and the $q^2$-ary optimal $(r,\delta)$-LRCs in Theorem \ref{Opt-1} share the same length, dimension, and minimum distance only when $m=q-1$; their parameters differ in all other cases. Therefore, the optimal $(r,\delta)$-LRCs constructed in Theorems \ref{Euclidean} and \ref{Opt-1} constitute two different families.
\end{remark}

\section{Infinite Families of Optimal quantum $(r,\delta)$-LRCs via Hermitian dual-containing MP codes}\label{section5}

In this section, we construct three infinite families of optimal quantum $(r,\delta)$-LRCs via Hermitian dual-containing MP codes.

\subsection{The first family of optimal quantum $(r,\delta)$-LRCs via Hermitian dual-containing MP codes}

\begin{thm}\label{MP-12}
Let $N|(q^2-1)$ and $N\nmid (q+1)$, where $q\geq 5$ is a prime power. Let $u,t\in \mathbb{N}^+$ and $v\in \mathbb{N}$ satisfy $v\leq u$ and $2u+v\leq q-2$. Then, there exists an infinite family of optimal quantum $(r,\delta)$-LRCs with parameters
\begin{align}
\big[\mspace{-3mu}\big[Nt(q-1),Nt(q-1)-2Ntu-2v, u+v+1\big]\mspace{-3mu}\big]_{q},\label{quan-fam12}
\end{align}
where $(r,\delta)=(q-1-u,u+1)$.
\end{thm}

\begin{IEEEproof}
Let $A$ be the $N\times N$ type II $\tau$-OD matrix from Lemma \ref{theorem20} for
\begin{align}\label{permu1}
\tau=\begin{pmatrix}
1 & 2 & 3 & \cdots  & N\\
t_{0}+1 & t_{1}+1& t_{2}+1 & \cdots  &  t_{N-1}+1
\end{pmatrix},
\end{align}
where $j+qt_{j}\equiv 0 \pmod{N}$ for $0\leq j \leq N-1$. Define $\mathcal{C}(A):=[\mathcal{C}_1,\ldots,\mathcal{C}_N]\cdot A$ with
$\mathcal{C}_1=\ldots=\mathcal{C}_{N-1}=\mathcal{C}(u,0,t)$ and $\mathcal{C}_N=\mathcal{C}(u,v,t)$.

We have $\tau(N)\neq N$. Since $2u+v\leq q-2$, it follows that $\mathcal{C}_{\tau(i)}^{\bot_{\mathrm{H}}}\subseteq \mathcal{C}_i$ for every $i\in [N]\backslash \{N,\tau(N)\}$. Moreover, $\mathcal{C}_{\tau(N)}^{\bot_{\mathrm{H}}}\subseteq \mathcal{C}_N$.
Therefore, $\mathcal{C}_{\tau(i)}^{\bot_{\mathrm{H}}}\subseteq \mathcal{C}_i$ for every $i\in[N]$.

Set $k_1=t(q-1)-tu$, $k_N=t(q-1)-tu-v$ and $r=q-1-u$. Since $2u+v\leq q-2$, we have
\begin{align*}
\bigg\lceil \frac{(N-1)k_1+k_N}{r}\bigg\rceil=\bigg\lceil \frac{Nt(q-1-u)-v}{q-1-u}\bigg\rceil=Nt
\end{align*}
and
\begin{align*}
(N-1)\bigg\lceil \frac{k_1}{r}\bigg\rceil+\bigg\lceil\frac{k_N}{r}\bigg\rceil=Nt.
\end{align*}
Therefore, $\Big\lceil \frac{(N-1)k_1+k_N}{r}\Big\rceil=(N-1)\big\lceil \frac{k_1}{r}\big\rceil+\big\lceil\frac{k_N}{r}\big\rceil$.
Additionally, $v\leq u$ ensures $u+v+1\leq 2(u+1)$, i.e., $d_{N}\leq 2\delta$.

By Theorem \ref{Main-thm}, $\mathcal{C}(A)$ induces an optimal quantum $(r,\delta)$-LRC with parameters in Eq. \eqref{quan-fam12}.
\end{IEEEproof}

\begin{example}\label{quan-ex2}
Let $q=7$, $N=3$, $u=2$, $v=1$ and $t=2$ in Theorem \ref{MP-12}. Let $\omega_1$ and $\omega_2$ be primitive $6$-th roots of unity in $\mathbb{F}_{7^2}$.
Define $\mathcal{C}_1=\mathcal{C}_2=\mathcal{C}(2,0,2)$ with parity-check matrices
\begin{align*}
H_1=H_2=\begin{pmatrix}
1 & \omega_1 & \cdots&  \omega_1^{5}& 0& 0 &\cdots&0\\
1 & \omega_1^2 & \cdots&  \omega_1^{2\times 5}& 0& 0 & \cdots&0\\
0&  0& \cdots& 0&  1 & \omega_2 & \cdots&  \omega_2^{5}\\
0&  0& \cdots& 0&  1 & \omega_2^2 & \cdots&  \omega_2^{2\times 5}
\end{pmatrix},
\end{align*}
and define $\mathcal{C}_3=\mathcal{C}(2,1,2)$ with parity-check matrix
\begin{align*}
H_3=\begin{pmatrix}
1 & \omega_1 & \cdots&  \omega_1^{5}& 0& 0 &\cdots&0\\
1 & \omega_1^2 & \cdots&  \omega_1^{2\times 5}& 0& 0 & \cdots&0\\
0&  0& \cdots& 0&  1 & \omega_2 & \cdots&  \omega_2^{5}\\
0&  0& \cdots& 0&  1 & \omega_2^2 & \cdots&  \omega_2^{2\times 5}\\
 1 & \omega_1^{3} & \cdots&  \omega_1^{3\times 5} & 1 & \omega_2^{3} & \cdots&  \omega_2^{3\times 5}
\end{pmatrix}.
\end{align*}
Let $A$ be a $\tau$-OD matrix of size $3\times 3$ with $\tau=(2,3)$ from Lemma \ref{theorem20}. By Theorem \ref{MP-12}, $\mathcal{C}(A):=[\mathcal{C}_1,\mathcal{C}_2,\mathcal{C}_3]\cdot A$ induces an optimal quantum $(4,3)$-LRC with parameters $[\mspace{-2mu}[36,10,4]\mspace{-2mu}]_{11}$.
\end{example}

\subsection{The second family of optimal quantum $(r,\delta)$-LRCs from Hermitian dual-containing MP codes}

\begin{thm}\label{MP-22}
Let $N\mid(q^2-1)$ and $N\nmid (q+1)$, where $q\geq 8$ is a prime power. Let $s,v,t\in \mathbb{N}^+$ and $l\in \mathbb{N}$ satisfy
\begin{itemize}
\item  $v|(q-1)$, $2s+l\leq v-1$, $l \leq \lfloor \frac{s}{2}\rfloor +1$;

\item  $Nvt\leq q+v-s-l-2$.
\end{itemize}
Then, there exists an infinite family of optimal quantum $(r,\delta)$-LRCs with parameters
\begin{align}
\big[\mspace{-3mu}\big[Nt(2q+v-2),Nt(2q+v-2)-4Nst-2l, s+l+1\big]\mspace{-3mu}\big]_{q}, \label{quan-fam22}
\end{align}
where $(r,\delta)=(q+v-s-1,s+1)$.
\end{thm}

\begin{IEEEproof}
Define $\mathcal{C}(A):=[\mathcal{C}_1,\ldots,\mathcal{C}_N]\cdot A$, where $A$ is the $N\times N$ $\tau$-OD matrix with
$\tau$ being defined in Eq. \eqref{permu1}, $\mathcal{C}_1=\ldots=\mathcal{C}_{N-1}=\mathcal{C}(s,0,v,t)$ and $\mathcal{C}_N=\mathcal{C}(s,l,v,t)$.

We have $\tau(N)\neq N$. Since $2s+l\leq v-1$, it follows that $\mathcal{C}_{\tau(i)}^{\bot_{\mathrm{H}}}\subseteq \mathcal{C}_i$ for every $i\in [N]\backslash \{N,\tau(N)\}$. Moreover, $\mathcal{C}_{\tau(N)}^{\bot_{\mathrm{H}}}\subseteq \mathcal{C}_N$. Therefore, $\mathcal{C}_{\tau(i)}^{\bot_{\mathrm{H}}}\subseteq \mathcal{C}_i$ for every $i\in[N]$.

Set $k_1=t(2q+v-2)-2st$, $k_N=t(2q+v-2)-2st-l$, $r=q+v-s-1$ and $\delta=s+1$. Since $Nvt\leq q+v-s-l-2$, we have
\begin{align*}
\bigg\lceil \frac{(N-1)k_1+k_N}{r}\bigg\rceil=\bigg\lceil \frac{2Nt(q+v-s-1)-Ntv-l}{q+v-s-1}\bigg\rceil=2Nt
\end{align*}
and
\begin{align*}
(N-1)\bigg\lceil \frac{k_1}{r}\bigg\rceil+\bigg\lceil\frac{k_N}{r}\bigg\rceil=2Nt.
\end{align*}
Therefore, $\Big\lceil \frac{(N-1)k_1+k_N}{r}\Big\rceil=(N-1)\big\lceil \frac{k_1}{r}\big\rceil+\big\lceil\frac{k_N}{r}\big\rceil$.
Additionally, $l \leq \lfloor \frac{s}{2}\rfloor +1$ ensures $s+l+1\leq 2(s+1)$, i.e., $d_{N}\leq 2\delta$.

By Theorem \ref{Main-thm}, $\mathcal{C}(A)$ induces an optimal quantum $(r,\delta)$-LRC with parameters in Eq. \eqref{quan-fam22}.
\end{IEEEproof}

\begin{example}\label{quan-ex4}
Let $q=13$, $v=4$, $s=2$, $t=1$, $l=1$ and $N=3$ in Theorem \ref{MP-22}. Let $\omega$ be a primitive $12$-th root of unity and let $\zeta$ be a primitive $4$-th root of unity. Define $x_i=\lambda \omega^{i}$, $z_i=\mu \omega^{i}$ and $y_j=\zeta^{j}$ for $0\leq i\leq 11$ and $0\leq j\leq3$, where $\lambda=g$ and $\mu=g^2$, with $g$ being a primitive element of $\mathbb{F}_{13^{2}}$. Let $\mathcal{C}_1=\mathcal{C}_2=\mathcal{C}(2,0,4,1)$ be linear codes with parity-check matrices
\begin{align*}
H_1=H_2=\begin{pmatrix}
x_0 & x_1 & \cdots& x_{11}&y_0 &y_1&y_2& y_3& 0& 0 & \cdots& 0\\
x_0^2 & x_1^2 & \cdots& x_{11}^2&y_0^2 &y_1^2&y_2^2& y_{3}^2& 0& 0 & \cdots& 0\\
0&0 & \cdots&0 &y_0 &y_1&y_2& y_3&z_0 & z_1 & \cdots& z_{11}\\
0&0 & \cdots&0 &y_0^2 &y_1^2&y_2^2& y_3^2&z_0^2 & z_1^2 & \cdots& z_{11}^2
\end{pmatrix},
\end{align*}
and let $\mathcal{C}_3=\mathcal{C}(2,1,4,1)$ be a linear code with parity-check matrix
\begin{align*}
H_3=\begin{pmatrix}
x_0 & x_1 & \cdots& x_{11}&y_0 &y_1&y_2& y_3& 0& 0 & \cdots& 0\\
x_0^2 & x_1^2 & \cdots& x_{11}^2&y_0^2 &y_1^2&y_2^2& y_{3}^2& 0& 0 & \cdots& 0\\
0&0 & \cdots&0 &y_0 &y_1&y_2& y_3&z_0 & z_1 & \cdots& z_{11}\\
0&0 & \cdots&0 &y_0^2 &y_1^2&y_2^2& y_3^2&z_0^2 & z_1^2 & \cdots& z_{11}^2\\
x_0^3 & x_1^3 & \cdots& x_{11}^3&y_0^3 &y_1^3&y_2^3& y_3^3&z_0^3 & z_1^3 & \cdots& z_{11}^3
\end{pmatrix}.
\end{align*}
Let $A$ be a $3\times 3$ type II $\tau$-OD matrix from Lemma \ref{theorem20} for $\tau=(2,3)$. By Theorem \ref{MP-22}, the MP code $\mathcal{C}(A):=[\mathcal{C}_1,\mathcal{C}_2,\mathcal{C}_3]\cdot A$ induces an optimal quantum $(14,3)$-LRC with parameters $[\mspace{-2mu}[84,58,4]\mspace{-2mu}]_{13}$.
\end{example}

\subsection{The third family of optimal quantum $(r,\delta)$-LRCs from Hermitian dual-containing MP codes}

In the following theorem, we give an infinite family of optimal quantum $(r,\delta)$-LRCs from MP codes with non-nested constituent codes.

\begin{thm}\label{MP-ess4}
Let $q\geq 5$ be an odd prime power. Let $u,t\in \mathbb{N}^+$ and $v\in \mathbb{N}$ satisfy $v\leq u$ and $u+v\leq \frac{q-3}{2}$.
Then, there exists an infinite family of optimal quantum $(r,\delta)$-LRCs with parameters
\begin{align}\label{quan-ess3}
\big[\mspace{-3mu}\big[2t(q-1),2t(q-1)-4tu-2v, u+v+1\big]\mspace{-3mu}\big]_{q},
\end{align}
where $(r,\delta)=(q-1-u,u+1)$.
\end{thm}

\begin{IEEEproof}
Let $\mathcal{C}_1=\mathcal{C}(u,0,t)$ and $\mathcal{C}_2=\mathcal{C}(u,v,t)$. Then,
\begin{itemize}
\item $\mathcal{C}_1$ is a $(q-1-u,u+1)$-LRC with parameters $[t(q-1),t(q-1)-tu,u+1]_{q^2}$;

\item $\mathcal{C}_2\subseteq \mathcal{C}_1$ is a $(q-1-u,u+1)$-LRC with parameters $[t(q-1),t(q-1)-tu-v,u+v+1]_{q^2}$.
\end{itemize}

Define $\mathcal{C}(A):=[\mathcal{C}_1,\mathcal{C}_2']\cdot A$, where
\begin{align*}
A=\begin{pmatrix}
1 & 1 \\
1 &-1
\end{pmatrix}.
\end{align*}

Since $u+v\leq \lfloor \frac{q-1}{2}\rfloor -1$, both $\mathcal{C}_1$ and $\mathcal{C}_2$ are Hermitian dual-containing.
Note that $d_{1}=\delta=u+1$. Besides, $v\leq u$ ensures $d_{2}\leq 2d_{1}$. The condition $u+v\leq\big\lfloor\frac{q-1}{2}\big\rfloor-1$ implies
\begin{align*}
\bigg\lceil \frac{k_1+k_2}{r}\bigg\rceil=\bigg\lceil \frac{2t(q-1-u)-v}{q-1-u}\bigg\rceil=2t
\end{align*}
and
\begin{align*}
\bigg\lceil \frac{k_1}{r}\bigg\rceil+\bigg\lceil\frac{k_2}{r}\bigg\rceil=2t.
\end{align*}
Therefore, $\lceil \frac{k_1+k_2}{r}\rceil=\lceil \frac{k_1}{r}\rceil +\lceil \frac{k_2}{r}\rceil=2t$.

By Theorem \ref{MP-ess2}, $\mathcal{C}(A)$ induces an optimal quantum $(r,\delta)$-LRC with parameters in Eq. \eqref{quan-ess3}.
\end{IEEEproof}

\begin{example}
Let $q=11$, $u=3$, $v=1$ and $t=2$ in Theorem \ref{MP-ess4}. Let $\omega_1$ and $\omega_2$ be primitive $10$-th roots of unity in $\mathbb{F}_{11^{2}}$.
Let $\mathcal{C}_1=\mathcal{C}(3,0,2)$ be the linear code with parity-matrix
\begin{align*}
H_1=\begin{pmatrix}
1 & \omega_1 & \cdots&  \omega_1^{9}& 0 &0  &\cdots &0\\
1 & \omega_1^2 & \cdots&  \omega_1^{2\times 9} &0 &0  &\cdots &0 \\
1 & \omega_1^3 & \cdots&  \omega_1^{3\times 9} & 0 &0  &\cdots &0 \\
0&  0& \cdots & 0  & 1 & \omega_2 & \cdots&  \omega_2^{9}\\
0&  0& \cdots & 0  & 1 & \omega_2^2 & \cdots&  \omega_2^{2\times 9}\\
0&  0& \cdots & 0  & 1 & \omega_2^3 & \cdots&  \omega_2^{3\times 9}
\end{pmatrix},
\end{align*}
and let $\mathcal{C}_2'$ be the linear code with parity-check matrix
\begin{align*}
H_2'=\begin{pmatrix}
1 & -\omega_1 &\cdots & -\omega_1^{9}& 0 & 0 &\cdots & 0\\
1 & -\omega_1^2 &\cdots & -\omega_1^{2\times 9}& 0 & 0 &\cdots & 0\\
1 & -\omega_1^3 &\cdots & -\omega_1^{3\times 9}& 0 & 0 &\cdots & 0\\
0 & 0 &\cdots & 0&-1 & -\omega_2 &\cdots & -\omega_2^{9}\\
0 & 0 &\cdots & 0&-1 & -\omega_2^2 &\cdots & -\omega_2^{2\times 9}\\
0 & 0 &\cdots & 0&-1 & -\omega_2^3 &\cdots & -\omega_2^{3\times 9}\\
1 & -\omega_1^4&\cdots & -\omega_1^{4\times 9}&-1 & -\omega_2^4 &\cdots & -\omega_2^{4\times 9}
\end{pmatrix}.
\end{align*}
Let
\begin{align*}
A=\begin{pmatrix}
1 & 1 \\
1 &-1
\end{pmatrix}.
\end{align*}
By Theorem \ref{MP-ess4}, the MP code $\mathcal{C}(A):=[\mathcal{C}_1,\mathcal{C}_2']\cdot A$ induces an optimal quantum $(7,4)$-LRC with parameters $[\mspace{-2mu}[40,14,5]\mspace{-2mu}]_{11}$.
\end{example}

\begin{remark}\label{remark7}
Notably, one can verify that the constituent codes $\mathcal{C}_1$ and $\mathcal{C}_2'$ of the MP code $\mathcal{C}(A)=[\mathcal{C}_1,\mathcal{C}_2']\cdot A$ used to construct the optimal quantum $(r,\delta)$-LRCs in Theorem \ref{MP-ess4} are non-nested, i.e., $\mathcal{C}_2'\nsubseteq \mathcal{C}_1$.
Therefore, the optimal quantum $(r,\delta)$-LRCs in Theorem \ref{MP-ess4} is constructed from MP codes with non-nested constituent codes, distinguishing it from the constructions in Theorems \ref{MP-12} and \ref{MP-22}, which rely on MP codes with nested constituent codes.
\end{remark}

\section{Infinite Families of Optimal Quantum $(r,\delta)$-LRCs via Euclidean Dual-containing MP codes}\label{section6}

In this section, we first construct two new infinite families of type I $\tau$-OD matrices over $\mathbb{F}_{q}$.
Applying these matrices, we then present two new infinite families of optimal quantum $(r,\delta)$-LRCs via Euclidean dual-containing MP codes.

\subsection{Two infinite families of type I $\tau$-OD matrices}

\begin{thm}\label{infinite-type I}
Let $q=p^{e}$ be an odd prime power. Let $q-1=(N-1)s$ with $N\geq 3$ and $s\geq2$. Then, the following two statements hold.
\begin{itemize}
\item [(1)] If $p\mid N$, then there exists an infinite family of $N\times N$ type \textrm{I} $\tau$-OD matrices over $\mathbb{F}_{q}$, where
\begin{align*}
\tau=\prod_{i=1}^{\left\lceil\frac{N}{2}\right\rceil}(i,N+1-i).
\end{align*}

\item [(2)] If $p\nmid N$, then there exists an infinite family of $N\times N$ type \textrm{I} $\tau$-OD matrices over $\mathbb{F}_{q}$, where
\begin{align*}
\tau=\prod_{i=2}^{\left\lceil\frac{N}{2}\right\rceil}(i,N+1-i).
\end{align*}
\end{itemize}
\end{thm}

\begin{IEEEproof}
Let $g$ be a primitive element of $\mathbb{F}_{q}$. Define an $N\times N$ matrix over $\mathbb{F}_{q}$ as follows:
\begin{align}\label{equation22}
A=\begin{pmatrix}
1&1&1&1&\cdots&1\\
0&1&g^{s}&g^{2s}&\cdots&g^{(N-2)s}\\
0&1&g^{2s}&g^{4s}&\cdots&g^{2(N-2)s}\\
\vdots&\vdots&\vdots&\vdots&\ddots&\vdots\\
0&1&g^{(N-2)s}&g^{2(N-2)s}&\cdots&g^{(N-2)^{2}s}\\
0&1&1&1&\cdots&1\\
\end{pmatrix}.
\end{align}

It is straightforward to verify that $A$ is NSC, and
\begin{align*}
AA^{\top}=\begin{pmatrix}
N&0&\cdots&0&N-1\\
0&0&\cdots&N-1&0\\
\vdots&\vdots&\begin{rotate}{65}$\ddots$\end{rotate}&\vdots&\vdots\\
0&N-1&\cdots&0&0\\
N-1&0&\cdots&0&N-1\\
\end{pmatrix}.
\end{align*}

By Lemma \ref{LA} and Theorem \ref{theorem13}, statements (1) and (2) hold. Therefore, we complete the whole proof.
\end{IEEEproof}

\begin{remark}
The construction of infinite families of $\tau$-OD matrices is an interesting and challenging problem, as proposed in \cite[Sect. 5]{Cao2025Entanglement}. Theorem \ref{infinite-type I} illustrates that there exist two infinite families of type I $\tau$-OD matrices, which, to the best of our knowledge, represent two novel infinite families.
\end{remark}

\subsection{The first family of optimal quantum $(r,\delta)$-LRCs from Euclidean dual-containing MP codes}

\begin{thm}\label{MP-32}
Let $q-1=(N-1)s$, where $q=p^{e}\geq 5$ is an odd prime power, $N\geq 3$, $s\geq2$ and $p\mid N$. Let $u,t,m\in \mathbb{N}^+$ and $v\in \mathbb{N}$ satisfy $v\leq u$,
$2u+v\leq m-1$ and $m\mid(q-1)$. Then, there exists an infinite family of optimal quantum $(r,\delta)$-LRCs with parameters
\begin{align}\label{quan-fam32}
\big[\mspace{-3mu}\big[Ntm,Ntm-2Ntu-2v, u+v+1\big]\mspace{-3mu}\big]_{q},
\end{align}
where $(r,\delta)=(m-u,u+1)$.
\end{thm}

\begin{IEEEproof}
Let $A$ be an $N\times N$ type I $\tau$-OD matrix from Theorem \ref{infinite-type I} (1) for
\begin{align*}
\tau=\prod_{i=1}^{\left\lceil\frac{N}{2}\right\rceil}(i,N+1-i).
\end{align*}
Define $\mathcal{C}(A):=[\mathcal{C}_1,\ldots,\mathcal{C}_N]\cdot A$, with $\mathcal{C}_1=\ldots=\mathcal{C}_{N-1}=\mathcal{D}(u,0,t,m)$ and $\mathcal{C}_N=\mathcal{D}(u,v,t,m)$.

Note that $\tau(1)=N$ and $\tau(N)=1$. Since $2u+v\leq m-1$, we have $\mathcal{C}_{1}^{\bot_{\mathrm{E}}}\subseteq \mathcal{C}_N$ and, equivalently, $\mathcal{C}_{N}^{\bot_{\mathrm{E}}}\subseteq \mathcal{C}_1$. The condition $2u+v\leq m-1$ implies $\mathcal{C}_{\tau(i)}^{\bot_{\mathrm{E}}}\subseteq \mathcal{C}_i$ for every $i\in [N]\backslash \{1,N\}$. Therefore, $\mathcal{C}_{\tau(i)}^{\bot_{\mathrm{E}}}\subseteq \mathcal{C}_i$ for every $i\in [N]$.

Set $k_1=tm-tu$, $k_N=tm-tu-v$, $r=m-u$ and $\delta=u+1$. Since $2u+v\leq m-1$, we have
\begin{align*}
\bigg\lceil \frac{(N-1)k_1+k_N}{r}\bigg\rceil=\bigg\lceil\frac{Nt(m-u)-v}{m-u}\bigg\rceil=Nt
\end{align*}
and
\begin{align*}
(N-1)\bigg\lceil \frac{k_1}{r}\bigg\rceil+\bigg\lceil\frac{k_N}{r}\bigg\rceil=Nt.
\end{align*}
Therefore, $\Big\lceil \frac{(N-1)k_1+k_N}{r}\Big\rceil=(N-1)\big\lceil \frac{k_1}{r}\big\rceil+\big\lceil\frac{k_N}{r}\big\rceil$.
Additionally, $v\leq u$ ensures $u+v+1\leq 2(u+1)$, i.e., $d_{N}\leq 2\delta$.

By Theorem \ref{Main-thm}, $\mathcal{C}(A)$ induces an optimal quantum $(r,\delta)$-LRC with parameters in Eq. \eqref{quan-fam32}.
\end{IEEEproof}

\begin{example}\label{quan-ex2}
Let $q=9$, $u=2$, $t=2$, $v=1$, $N=3$ and $m=8$ in Theorem \ref{MP-32}. Let $\zeta_1$ and $\zeta_2$ be primitive $8$-th roots of unity in $\mathbb{F}_{9}$.
Let $\mathcal{C}_1=\mathcal{C}_2=\mathcal{D}(2,0,2,8)$ be linear codes with parity-check matrices
\begin{align*}
H_1=H_2=\begin{pmatrix}
1 & \zeta_1 & \cdots&  \zeta_1^{7}& 0& 0 &\cdots&0\\
1 & \zeta_1^2 & \cdots&  \zeta_1^{2\times 7}& 0& 0 & \cdots&0\\
0&  0& \cdots& 0&  1 & \zeta_2 & \cdots&  \zeta_2^{7}\\
0&  0& \cdots& 0&  1 & \zeta_2^2 & \cdots&  \zeta_2^{2\times 7}
\end{pmatrix},
\end{align*}
and let $\mathcal{C}_3=\mathcal{D}(2,1,2,8)$ be the linear code with parity-check matrix
\begin{align*}
H_3=\begin{pmatrix}
1 & \zeta_1 & \cdots&  \zeta_1^{7}& 0& 0 &\cdots&0\\
1 & \zeta_1^2 & \cdots&  \zeta_1^{2\times 7}& 0& 0 & \cdots&0\\
0&  0& \cdots& 0&  1 & \zeta_2 & \cdots&  \zeta_2^{7}\\
0&  0& \cdots& 0&  1 & \zeta_2^2 & \cdots&  \zeta_2^{2\times 7}\\
 1 & \zeta_1^{3} & \cdots&  \zeta_1^{3\times 7} & 1 & \zeta_2^{3} & \cdots&  \zeta_2^{3\times 7}
\end{pmatrix}.
\end{align*}
Let $A$ be a $3\times 3$ type I $\tau$-OD matrix from Theorem \ref{infinite-type I} (1) for $\tau=(1,3)$. By Theorem \ref{MP-32}, the MP code $\mathcal{C}(A):=[\mathcal{C}_1,\mathcal{C}_2,\mathcal{C}_3]\cdot A$ induces an optimal quantum $(6,3)$-LRC with parameters $[\mspace{-2mu}[48,22,4]\mspace{-2mu}]_{9}$.
\end{example}

\subsection{The second family of optimal quantum $(r,\delta)$-LRCs from Euclidean dual-containing MP codes}

\begin{thm}\label{MP-31}
Let $q-1=(N-1)s$, where $q=p^{e}\geq 5$ is an odd prime power, $N\geq 3$, $s\geq2$ and $p\nmid N$. Let $u,t,m\in \mathbb{N}^+$ and $v\in \mathbb{N}$ satisfy $v\leq u$,
$u+v\leq \lfloor \frac{m}{2}\rfloor -1$ and $m\mid(q-1)$. Then, there exists an infinite family of optimal quantum $(r,\delta)$-LRCs with parameters
\begin{align}\label{quan-fam31}
\big[\mspace{-3mu}\big[Ntm,Ntm-2Ntu-2v, u+v+1\big]\mspace{-3mu}\big]_{q},
\end{align}
where $(r,\delta)=(m-u,u+1)$.
\end{thm}

\begin{IEEEproof}
Let $A$ be an $N\times N$ type I $\tau$-OD matrix from Theorem \ref{infinite-type I} (2) for
\begin{align*}
\tau=\prod_{i=2}^{\left\lceil\frac{N}{2}\right\rceil}(i,N+1-i).
\end{align*}
Define $\mathcal{C}(A)=[\mathcal{C}_1,\ldots,\mathcal{C}_N]\cdot A$, with $\mathcal{C}_1=\ldots =\mathcal{C}_{N-1}=\mathcal{D}(u,0,t,m)$ and $\mathcal{C}_N=\mathcal{D}(u,v,t,m)$.

Note that $\tau(1)=1$ and $\tau(N)=N$. Since $u+v\leq \lfloor \frac{m}{2}\rfloor -1$, we have $\mathcal{C}_{1}^{\bot_{\mathrm{E}}}\subseteq \mathcal{C}_1$ and $\mathcal{C}_{N}^{\bot_{\mathrm{E}}}\subseteq \mathcal{C}_N$, respectively. Observe that $\mathcal{C}_{\tau(i)}^{\bot_{\mathrm{E}}}\subseteq \mathcal{C}_i$ for every $i\in [N]\backslash \{1,N\}$. Therefore, $\mathcal{C}_{\tau(i)}^{\bot_{\mathrm{E}}}\subseteq \mathcal{C}_i$ for every $i\in [N]$.

Set $k_1=tm-tu$, $k_N=tm-tu-v$, $r=m-u$ and $\delta=u+1$. Since  $u+v\leq \lfloor \frac{m}{2}\rfloor -1$, we have
\begin{align*}
\bigg\lceil \frac{(N-1)k_1+k_N}{r}\bigg\rceil=\bigg\lceil\frac{Nt(m-u)-v}{m-u}\bigg\rceil=Nt
\end{align*}
and
\begin{align*}
(N-1)\bigg\lceil \frac{k_1}{r}\bigg\rceil+\bigg\lceil\frac{k_N}{r}\bigg\rceil=Nt.
\end{align*}
Therefore, $\Big\lceil \frac{(N-1)k_1+k_N}{r}\Big\rceil=(N-1)\big\lceil \frac{k_1}{r}\big\rceil+\big\lceil\frac{k_N}{r}\big\rceil$.
Additionally, $v\leq u$ ensures $u+v+1\leq 2(u+1)$, i.e., $d_{N}\leq 2\delta$.

By Theorem \ref{Main-thm}, $\mathcal{C}(A)$ induces an optimal quantum $(r,\delta)$-LRC with parameters in Eq. \eqref{quan-fam31}.
\end{IEEEproof}

\begin{example}\label{quan-ex1}
Let $q=13$, $u=2$, $v=1$, $t=2$, $N=3$ and $m=12$. Let $\zeta_1$ and $\zeta_2$ be primitive $12$-th roots of unity in $\mathbb{F}_{13}$.
Let $\mathcal{C}_1=\mathcal{C}_2=\mathcal{D}(2,0,2,12)$ be linear codes with parity-check matrices
\begin{align*}
H_1=H_2=\begin{pmatrix}
1 & \zeta_1 & \cdots&  \zeta_1^{11}& 0 &0  &\cdots &0\\
1 & \zeta_1^2 & \cdots&  \zeta_1^{2\times 11} &0 &0  &\cdots &0 \\
0&  0& \cdots & 0  & 1 & \zeta_2 & \cdots&  \zeta_2^{11}\\
0&  0& \cdots & 0  & 1 & \zeta_2^2 & \cdots&  \zeta_2^{2\times 11}
\end{pmatrix},
\end{align*}
and let $\mathcal{C}_3=\mathcal{D}(2,1,2,12)$ be the linear code with parity-check matrix
\begin{align*}
H_3=\begin{pmatrix}
1 & \zeta_1 & \cdots&  \zeta_1^{11}& 0 &0  &\cdots &0\\
1 & \zeta_1^2 & \cdots&  \zeta_1^{2\times 11} &0 &0  &\cdots &0 \\
0&  0& \cdots & 0  & 1 & \zeta_2 & \cdots&  \zeta_2^{11}\\
0&  0& \cdots & 0  & 1 & \zeta_2^2 & \cdots&  \zeta_2^{2\times 11}\\
1 & \zeta_1^3 & \cdots&  \zeta_1^{3\times 11} & 1 & \zeta_2^3 & \cdots&  \zeta_2^{3\times 11}
\end{pmatrix}.
\end{align*}
Let $A$ be an $3\times 3$ type I $\tau$-OD matrix from Theorem \ref{infinite-type I} (2), where $\tau$ is an identity permutation. By Theorem \ref{MP-31}, the MP code $\mathcal{C}(A):=[\mathcal{C}_1,\mathcal{C}_2,\mathcal{C}_3]\cdot A$ induces an optimal quantum $(10,3)$-LRC with parameters $[\mspace{-2mu}[72,46,4]\mspace{-2mu}]_{13}$.
\end{example}

\section{Concluding remarks}\label{sec:5}
In this paper, we studied optimal quantum $(r,\delta)$-LRCs from matrix-product (MP) codes.
We established a necessary and sufficient condition for an MP code to be an optimal $(r,\delta)$-LRC.
Based on this, we presented a characterization for optimal quantum $(r,\delta)$-LRCs from MP codes with nested constituent codes, and also studied optimal quantum $(r,\delta)$-LRCs constructed from MP codes with non-nested constituent codes.
Through Hermitian dual-containing and Euclidean dual-containing MP codes, we presented five infinite families of optimal quantum $(r,\delta)$-LRCs with flexible parameters.

\end{document}